\documentclass[a4paper,11pt]{article}
\usepackage{jheppub} 
\usepackage{lineno,color,slashed,changes,color}
\usepackage[normalem]{ulem}

\title{Muon $(g-2)$ and Thermal WIMP DM in ${\rm U(1)}_{L_\mu - L_\tau} $ Models}

\author[a]{Seungwon Baek,}
\author[b,c, *]{Jongkuk Kim} 
\author[b]{and P. Ko}
\affiliation[a]{The Institute of Basic Science, Korea University, Anam-ro 145, Seoul 02841, Korea}
\affiliation[b]{School of physics, Korea Institute for Advanced Study (KIAS), Seoul 02455, Republic of Korea}
\affiliation[c]{Department of Physics, Chung-Ang University, Seoul 06974, Korea}
\emailAdd{sbaek1560@gmail.com}
\emailAdd{jkkim@kias.re.kr}
\emailAdd{pko@kias.re.kr}
\preprint{KIAS-P22019}
\note[*]{corresponding Author}

\abstract{
The ${\rm U(1)}_{L_\mu - L_\tau}$  model is anomaly-free with the Standard Model (SM) fermion content, and can make substantial contributions to the muon $(g-2)$ at the level of $\Delta a_\mu \sim O(10) \times 10^{-10}$ for $M_{Z'} \sim O(10-100)$ MeV and 
$g_X \sim (4 - 8) \times 10^{-4}$.
In this light $Z'$ region, it was claimed that the model can also incorporate thermal WIMP dark matter (DM) if $M_{\rm DM} \sim M_{Z'}/2$. 
This setup relies on DM particles annihilating into SM particles through a $Z'$-mediated $s$-channel.  
In this work, we show that this tight relationship between $M_{Z'}$ and $M_{\rm DM}$ 
can be evaded or nullified both for scalar and spin-1/2
DM by considering the contributions from the dark Higgs boson ($H_1$). 
The dark Higgs boson plays an important role, not only because it  gives mass 
to the dark photon but also because it introduces additional DM annihilation channels, 
including new final states such as $H_1 H_1$,  $Z' Z'$, and $Z' H_1$.
As a result, the model does not require a close mass correlation between the $Z'$ boson and dark matter 
$M_{\rm DM} \sim M_{Z'}/2$ any longer,  allowing for a broader range of mass possibilities for both scalar 
and fermionic dark matter types.   We explore in great details various  scenarios where the $U(1)$ symmetry 
is either fully broken or partially remains as discrete symmetries, $Z_2$ or $Z_3$.
This approach broadens the model's capacity to accommodate various WIMP dark matter phenomena in the light $Z'$ region where the muon $(g-2)_\mu$ makes a sensitive probe of the model.
}

\begin{document}
\maketitle
\flushbottom

\section{Introduction}
The anomalous magnetic moment of the muon, $a_\mu = (g-2)_\mu/2$, has been a testing ground for 
Quantum Electrodynamics (QED) in early days, and also served as a probe of Beyond the Standard Model 
(BSM) physics in various contexts.
The combined world average of measurements, including recent results announced by the Fermilab Muon $g-2$ Collaboration,  is given by~\cite{Muong-2:2023cdq}
\begin{equation}
a_\mu^{\rm exp} = (11\, 659\, 205.9 \pm 2.2) \times 10^{-10}.
\end{equation}

The SM predictions over the last half-century~\cite{Czarnecki:2002nt,Melnikov:2003xd,Davier:2010nc,Hagiwara:2011af,Aoyama:2012wk,Gnendiger:2013pva,Kurz:2014wya,Colangelo:2014qya,Masjuan:2017tvw,Colangelo:2017fiz,Davier:2017zfy,Keshavarzi:2018mgv,Hoferichter:2018kwz,Colangelo:2018mtw,Gerardin:2019vio,Hoferichter:2019mqg,Davier:2019can,Bijnens:2019ghy,Colangelo:2019uex,Keshavarzi:2019abf,Aoyama:2019ryr,Blum:2019ugy} have been summarized 
by the Muon $g-2$ Theory Initiative in their white paper~\cite{Aoyama:2020ynm}. 
However, as it is now outdated and does not reflect recent results from lattice collaborations and $e^+e^-$ experiments, 
we do not quote their prediction for the muon $(g-2)$.

In the SM prediction for the muon ($g-2$), the main theoretical uncertainty stems 
from the hadronic vacuum polarization (HVP), which has been traditionally derived 
from low-energy $e^+ e^- \to$ hadrons data via the dispersive method. 
Independent of traditional dispersive approaches to the HVP, there have been developments in the lattice QCD community to evaluate the HVP using nonperturbative lattice QCD 
methods.  In the year 2020, the BMW collaboration announced their results, 
\begin{equation}
\Delta a_\mu = (10.7 \pm 6.9) \times 10^{-10},
\label{eq:Delta_BMW}
\end{equation}
indicating a notably lesser  discrepancy of $1.5\sigma$~\cite{Borsanyi:2020mff} compared to the experimental data at that time. Afterwards, the BMW results have been further supported by additional lattice 
studies~\cite{Kuberski:2023qgx}.  Also an enhanced HVP contribution from the new measurements of 
$e^+ e^- \to \pi^+ \pi^-$ cross-section in the range 0.3 GeV $ < \sqrt{s} < 1.2$ GeV 
by CMD-3 collaboration~\cite{CMD-3:2023alj,CMD-3:2023rfe}  resulted in 
\begin{equation}
\Delta a_\mu = \left(4.9 \pm 5.5 \right) \times 10^{-10}, 
\label{eq:Delta_CMD3}
\end{equation}
reducing the discrepancy to within 1$\sigma$. 

{\it Given the many possibilities to treat/combine the dispersive and the 
Lattice QCD results on the HVP, we do not commit ourselves to a specific value 
of $\Delta a_\mu$ in this paper. 
Instead we will consider BSM models where the muon 
$(g-2)$ can be as large as $\Delta a_\mu \sim 10 \times 10^{-10}$, 
which is consistent with the BMW and CMD-3 results at 1$\sigma$, 
thus making the most sensitive probe of such BSM models.}   
A simple way to generate $\Delta a_\mu \sim 10 \times 10^{-10}$, if any, is to use 
a $\mathcal{O}(10^{1\pm 1} )$ MeV-scale new gauge boson, which couples to the muon with strength $\sim \mathcal{O}(10^{-4})$. 
Additionally, we consider one of the benchmark points where no discrepancy exists between experimental results and SM predictions for the muon $(g-2)$, even when accounting for the error bars from prospective experiments and theoretical predictions.
If this new particle also couples to dark matter (DM), it can affect the DM annihilation into the SM particles in the early universe and can influence the thermal relic density 
of DM. It can also act like dark radiation if it is light enough and in thermal equilibrium with the SM sector in the early universe.
Thus the light new gauge boson of $\mathcal{O}(10^{1\pm 1} )$ MeV 
can play important roles in particle physics and cosmology through the muon 
($g-2$), dark matter and dark radiation as well as the Hubble tension,
which makes the topic very interesting and worthwhile for detailed study.

In order to fully address these issues, we will choose three benchmark points for $\Delta a_\mu = 15.3 \times 10^{-10}, 12.5\times 10^{-10}$ and $\Delta a_\mu = 9.9 \times 10^{-13} \simeq 0$,
partly guided by the current status of the $\Delta a_\mu$ as described in the previous paragraphs (see Tab.~\ref{tab:1} and related discussions in Sec.~\ref{Sec:2} for 
the reasons behind choosing these three benchmark points),  and will study the allowed mass range of WIMP dark matter for each benchmark point.    
The main conclusions of our paper on dark matter phenomenology,  especially 
the dark Higgs boson being able to nullify the tight correlation 
$m_{\rm DM} \sim M_{Z'}/2$ obtained in earlier literature, will not change 
regardless of the specific value for $\Delta a_\mu$ in future analyses.

When adding an extra gauge boson that couples to the muon, it is important to maintain 
anomaly cancellations.  One of the simplest gauge groups that fulfills these conditions with 
the SM fermion contents is $U(1)_{L_\mu - L_\tau}$ \cite{He:1990pn, He:1991qd} .  
There have been comprehensive studies on the phenomenology of the $U(1)_{L_\mu - L_\tau}$ 
gauge extension of the SM  in the context of $\Delta a_\mu$ \cite{Baek:2001kca,Ma:2001md,Altmannshofer:2014pba,Park:2015gdo,Altmannshofer:2016oaq, Biswas:2016yan,  Huang:2021nkl,   Chen:2021vzk,  Zhou:2021vnf,Buras:2021btx, Borah:2021mri, Costa:2022oaa,Ho:2024cwk}, the lepton universality test \cite{Chun:2018ibr}, and leptophilic DM \cite{Baek:2008nz,Patra:2016shz,Zu:2021odn} for the PAMELA $e^+$ excess \cite{PAMELA:2008gwm}. 
One can also address some $B$ anomalies if one introduces more particles with  
nonzero ${\rm U(1)}_{L_\mu - L_\tau}$ charges
\cite{Altmannshofer:2014cfa, Crivellin:2015mga, Crivellin:2015lwa,Ko:2017yrd, Chen:2017usq,Baek:2017sew,Baek:2018aru,Baek:2019qte, Biswas:2019twf,Han:2019diw,Kumar:2020web,Chao:2021qxq, Borah:2021khc,Ko:2021lpx}. 

In this work, we explore the implications of the $\Delta a_\mu$ contribution and 
thermal weakly interacting massive particle (WIMP) DM within the framework of the 
$U(1)_{L_\mu - L_\tau}$ model in the light $Z'$ regime, $O(10)$ MeV $\lesssim m_{Z'} \lesssim O(100)$ MeV,
where the model can generate $\Delta a_\mu \sim O(10) \times 10^{-10}$.
It is widely acknowledged that this model can successfully account for thermal DM in the light $Z'$ region under two scenarios:  (i) when $M_{\rm DM} \sim M_{Z'}/2$, with dark matter mass in the range of 
$\mathcal{O}(10-100)$ MeV, a configuration which is very predictive as highlighted in previous studies \cite{Foldenauer:2018zrz,Holst:2021lzm,Drees:2021rsg}, 
or (ii) by considering dark matter with a significantly large $U(1)$ charge \cite{Asai:2020qlp}.

This paper proposes an alternative perspective by considering the inclusion of the dark Higgs boson, which emerges from the spontaneous symmetry breaking mechanism that gives mass to the dark photon $Z'$ through a nonzero vacuum expectation value (VEV) $\langle \Phi \rangle =v_\Phi/\sqrt{2} \neq 0$. The influence of the dark Higgs boson on dark matter physics has been largely overlooked in existing literature. Nevertheless, the present authors have contributed to a series of studies addressing this gap, examining various phenomena such as the galactic center $\gamma$-ray excess~\cite{Ko:2014gha,Ko:2014loa,Baek:2014kna,Ko:2015ioa}, the formation of dark matter bound states and their impact on relic density calculations~\cite{Ko:2019wxq}, and dark matter searches at high-energy colliders, including the International Linear Collider (ILC)~\cite{Ko:2016xwd,Kamon:2017yfx}, the Large Hadron Collider (LHC), and future 100 TeV $pp$ colliders~\cite{Baek:2015lna,Ko:2016ybp,Dutta:2017sod,Ko:2018mew}, the role of $p$-wave annihilations in light dark matter scenarios which helps circumvent stringent constraints from big bang nucleosynthesis (BBN) and the cosmic microwave background (CMB)~\cite{Baek:2020owl}, as well as the 
Higgs-portal assisted Higgs inflation scenarios~\cite{Kim:2014kok,Khan:2023uii} 
(for reviews on these topics and phenomenological constraints on the dark Higgs boson, see references \cite{Ko:2016yfb,Ko:2018qxz,Clarke:2013aya,Ibe:2021fed}).

Our key assertion is that the inclusion of the dark Higgs boson can lead to substantial changes in dark matter phenomenology. In this paper, we demonstrate that by incorporating the dark Higgs boson, it is possible to entirely bypass the conventional $s$-channel annihilation condition, 
\begin{eqnarray}
M_{\rm DM} \sim M_{Z'}/2,	\label{eq:res}
\end{eqnarray}
which is typically required for achieving the correct dark matter thermal relic density.

\section{The ${\rm U}(1)_{L_\mu - L_\tau}$ Models} \label{Sec:2}

The ${\rm U(1)}_{L_\mu - L_\tau}$, which will also be referred to as ${\rm U(1)}_X$ henceforth, extensions of the SM are anomaly-free, 
not requiring the introduction of additional chiral fermions~\cite{He:1990pn, He:1991qd}. In the minimal setup accommodating thermal WIMP DM, one introduces a dark Higgs $\Phi$, 
along with scalar or fermion dark matter candidates, $X$ and $\chi$, respectively. We assume both the $\Phi$ and DM are singlets under the SM gauge group
${\rm SU}(3)_C \times {\rm SU}(2)_L \times {\rm U(1)}_Y$.
The charge assignments under 
the ${\rm U(1)}_X$ symmetry for these particles are specified as follows:
\begin{equation}
Q_X ( \mu , \nu_\mu , \tau , \nu_\tau , X, \chi , \Phi ) = (1,1,-1,-1,Q_X,Q_\chi, Q_\Phi ).
\end{equation}
Specific assignments of the DM and dark Higgs charges will be discussed in later sections.

The model Lagrangian is dictated by gauge invariance and is given by
\begin{eqnarray}
\mathcal{L} &=& \mathcal{L}_{SM} -\frac{1}{4}Z'^{\mu\nu}Z'_{\mu\nu} - g_X Z'_\mu 
\left( \bar{\ell}_\mu \gamma^\mu \ell_\mu - \bar{\ell}_\tau \gamma^\mu \ell_\tau +\bar{\mu}_R \gamma^\mu \mu_R - \bar{\tau}_R \gamma ^\mu \tau_R \right)\\
& + & D_\mu \Phi^\dagger D^\mu \Phi - \lambda_\Phi \left( \Phi^\dagger \Phi - \frac{v_\Phi^2}{2} \right)^2 - \lambda_{\Phi H} \left( \Phi^\dagger \Phi - \frac{v_\Phi^2}{2} \right) \left( H^\dagger H  - \frac{v^2}{2} \right)  + \mathcal{L}_{\rm DM} ,    \nonumber
\end{eqnarray}
where $g_X$ is the ${\rm U(1)}_X $ gauge coupling, and $D_\mu \Phi = ( \partial_\mu + ig_X Q_\Phi Z'_\mu ) \Phi$ is the covariant derivative acting on the dark Higgs field. 
The term $ \mathcal{L}_{\rm DM}$ will depend on the spin and the ${\rm U(1)}_X$ charge of the dark matter, and will be specified case by case in the later sections.

The phase where the $U(1)$ symmetry is spontaneously broken is described by
\[
\Phi (x) = \frac{1}{\sqrt{2}} \left( v_\Phi + \phi (x) \right) ,
\]
where $v_\Phi$ is the vacuum expectation value (VEV) of the dark Higgs field\added{,} and we assume $v_\Phi>0$. 
The mass of  {\it the dark photon} $Z'$ is generated by the nonzero VEV of $\Phi$:
\begin{eqnarray}
M_{Z'} &=& g_X \vert Q_\Phi \vert v_\Phi.
\end{eqnarray}

In this model, there are  two CP-even neutral scalar bosons, 
which mix with each other through the Higgs portal coupling, $\lambda_{\Phi H}$.
We define the mixing matrix $O$ connecting the interaction and mass eigenstates as follows:
\begin{equation}
\left(
\begin{array}{c}
\phi \\ h
\end{array}
\right)
= O 
\left(
\begin{array}{c}
H_1 \\ H_2
\end{array}
\right)
\equiv  
\left(
\begin{array}{cc}
c_\alpha & s_\alpha \\
-s_\alpha & c_\alpha 
\end{array}
\right)
\left(
\begin{array}{c}
H_1 \\ H_2
\end{array}
\right),
\end{equation}
where $s_\alpha (c_\alpha) \equiv \sin\alpha (\cos\alpha)$, and $(\phi, h)$ and $(H_1, H_2)$ are the interaction and mass eigenstates with masses 
$M_{H_i}$ ($i=1,2$), respectively. 
The mixing angle $\alpha$ is defined by
\begin{eqnarray}
\tan 2\alpha = \frac{\lambda_{\Phi H} v_\Phi v_H}{\lambda_H v^2_H -\lambda_\Phi v^2_\Phi } ,
\end{eqnarray}
where $v_H \simeq 246$ GeV is the VEV of the SM Higgs.  
The mass matrix in the interaction basis $(\phi, h)$ can be expressed in terms of the 
physical parameters as follows:
\begin{equation}
\left(
\begin{array}{cc}
2 \lambda_\Phi v_\Phi^2 & \lambda_{\Phi H} v_\Phi v_H \\ 
\lambda_{\Phi H} v_\Phi v_H & 2 \lambda_H v_H^2
\end{array}
\right) =
\left(
\begin{array}{cc}
M_{H_1}^2 {c_\alpha^2} + M_{H_2}^2 {s_\alpha^2} & (M_{H_2}^2 -M_{H_1}^2 ) c_\alpha s_\alpha \\ 
(M_{H_2}^2 -M_{H_1}^2 ) c_\alpha s_\alpha & M_{H_1}^2 {s_\alpha^2} + M_{H_2}^2 {c_\alpha^2}
\end{array}
\right). \label{eq:mass_matrix}
\end{equation}
Now we can consider $M_{H_i}$ and $\sin\alpha$ as independent parameters.
	
The mixing angle $\alpha$ is constrained by Higgs decay data and is expected to be rather small, with $\sin\alpha \lesssim 10^{-3}$ ($10^{-2}$) for the benchmark points we consider (see below for the definition of the benchmark points). Consequently, $H_1$ ($H_2$) is predominantly a dark-Higgs-like (SM Higgs-like) state. We set $M_{H_2} = 125.25$ GeV. Given our interest in the light dark Higgs scenario ($M_{H_1} < M_{H_2}$) within our model, we can consistently assume $|\alpha| < \pi/4$. In the limit of a small mixing angle, it follows that $H_1 \simeq \phi$ and $H_2 \simeq h$.
	
The gauge interaction terms between $Z'$ and SM particles are given by
\begin{align}
\mathcal{L}_{Z'} \supset -g_X  Z'_\mu \left( \bar{\mu}\gamma^\mu \mu -\bar{\tau}\gamma^\mu \tau  +\bar{\nu}_{L_\mu} \gamma^\mu \nu_{L_ \mu} -\bar{\nu}_{L_\tau} \gamma^\mu \nu_{L_\tau} \right). 
\end{align}
After spontaneous symmetry breaking, the  Higgs-gauge and Higgs-self interactions in terms of $\phi$ and $h$ are given by 
\begin{align}
\mathcal{L}_{\phi} &\supset \frac{1}{2}g^2_X Q^2_\Phi Z'^\mu Z'_\mu \phi^2 + g^2_X Q^2_\Phi v_\Phi Z'^\mu Z'_\mu \phi  -\lambda_\Phi v_\Phi \phi^3 - \lambda_H v_H h^3 -\frac{\lambda_{\Phi H}}{2} v_\Phi \phi h^2 - \frac{\lambda_{\Phi H}}{2} v_H \phi^2 h.
\end{align}

We assume that the kinetic mixing between the $Z'$ and the hypercharge gauge boson $B$ is zero at a high scale. However, small mixings of the $Z'$ boson with the SM photon $A$ and the $Z$ boson are generated radiatively:
\begin{equation}
\mathcal{L}_{\epsilon} = -\frac{\epsilon_A}{2} F_{\mu\nu}Z'^{\mu\nu} -\frac{\epsilon_Z}{2} Z_{\mu\nu}Z'^{\mu\nu},
\end{equation}
where $F_{\mu\nu}$ and $Z_{\mu\nu}$ represent the field strengths for the SM photon and $Z$ boson, respectively. The canonical basis of gauge fields can be recovered by the transformation:
\begin{eqnarray}
A_\mu &\to& A_\mu +\epsilon_A Z'_\mu, \nonumber\\
Z_\mu &\to& Z_\mu -\epsilon_Z \frac{M^2_{Z'}}{M^2_Z} Z'_\mu, \nonumber\\
Z'_\mu &\to& Z'_\mu +\epsilon_Z Z_\mu.
\end{eqnarray}
The interaction terms resulting from the mixing are given by:
\begin{equation}	
\mathcal{L}_{\epsilon} = - \epsilon_A Z'_\mu J^\mu_{\text{em}} + \frac{g}{\cos\theta_W}\epsilon_Z \frac{M^2_{Z'}}{M^2_Z} Z'_\mu J^\mu_{\text{NC}},
\end{equation}
where $J^\mu_{\text{em}}$ and $J^\mu_{\text{NC}}$ are the electromagnetic and neutral currents, respectively.
At the one-loop level, the mixing parameters $\epsilon_A$ and $\epsilon_Z$ are obtained by calculating the diagrams with $\mu$ and $\tau$ running inside the loop:
\begin{equation}
\epsilon_A = -\frac{e g_X}{12\pi^2} \ln\left( \frac{M^2_\tau}{M^2_\mu} \right) \simeq -\frac{g_X}{70}.
\end{equation}
The parameter $\epsilon_Z$ is further suppressed by $M^2_{Z'}/M^2_Z$ , and thus becomes negligible for the light $Z'$ mass region of interest.

\begin{figure}
\includegraphics[width=0.7\linewidth]{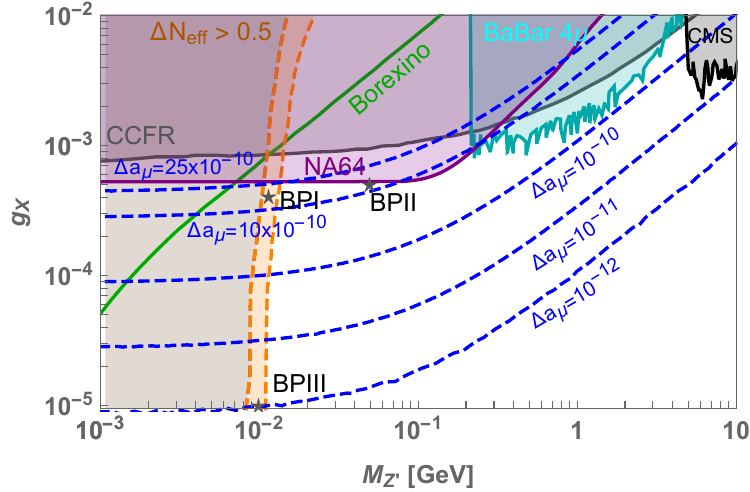}
\centering
\hspace{0.5cm}
\caption{
The blue dashed lines represent the additional contributions from the light $Z'$ boson. The cyan, black, purple and brown regions are excluded by other experimental bounds. The region above the green solid line is ruled out by the Borexino experiment. 
The area within the narrow orange strip can resolve the Hubble tension. We consider three benchmark points, which are defined in Tab.~\ref{tab:1}.
}\label{LmuLtau}
\end{figure}

In this model, the contribution of the $Z'$ boson to $\Delta a_\mu$ at one-loop is expressed by the following integral \cite{Baek:2001kca, Ma:2001md, Baek:2008nz, Altmannshofer:2014pba}:
\begin{equation}
\Delta a_\mu = \frac{\alpha_X}{2 \pi} \int_0^1 dx \frac{2 M_\mu^2 x^2 (1-x)}{x^2 M_\mu^2 + (1-x) M_{Z'}^2}\replaced{,}{.}	\label{eq:a_mu_Zprime}
\end{equation}
where $\alpha_X=g_X^2/4\pi$, with $g_X$ being the $U(1)_X$ gauge coupling constant. Considering $g_X \sim (4-8) \times 10^{-4}$, Eq.~(\ref{eq:a_mu_Zprime}) generates $\Delta a_\mu \sim 10 \times 10^{-10}$ in the limit $M_{Z'} < M_\mu$.

Moreover, the $Z'$ boson can contribute to neutrino trident events during muon-neutrino scattering off a nucleus $N$, specifically in the process $\nu_\mu N \rightarrow \nu_\mu N \mu^+ \mu^-$ \cite{Altmannshofer:2014cfa}. Experimental data from the CHARM-II \cite{CHARM-II:1990dvf} and CCFR \cite{CCFR:1991lpl} collaborations impose a stringent constraint, excluding the parameter region with $M_{Z'} > O(1)$ GeV \cite{Altmannshofer:2014pba}\footnote{
However, if one extends the particle content, $U(1)_{L_\mu - L_\tau}$ models can easily accommodate $\Delta a_\mu$ at one loop levels involving new particles, even 
for the case of a heavy $Z'$.   See Refs.~\cite{Altmannshofer:2016oaq, Patra:2016shz, Huang:2021nkl, Zu:2021odn,  Chen:2021vzk,Borah:2021jzu, Singirala:2021gok, Zhou:2021vnf,Buras:2021btx, Qi:2021rhh, Borah:2021mri} for explicit model constructions in this direction.}.
In our study, we impose $2\sigma$ exclusion limit from the CCFR data. 

Constraints on the $(M_{Z'}, g_X)$-plane are also considered from searches at BaBar and the LHC for the $4 \mu$ channel, Borexino neutrino oscillation data, 
and measurements of $\Delta N_{\rm eff}$, as well as the NA64, as summarized in the Appendix. After accounting for all 
experimental bounds, the remaining parameter space for $\Delta a_\mu$ in the case of a light $Z'$ 
is illustrated in Fig.~\ref{LmuLtau}.  
The additional contributions from the light $Z'$ boson are represented by the blue dashed lines. 
Note that this parameter region is more sensitive to the muon ($g-2$) than other experiments, and
makes a nice playground to study $\Delta a_\mu$ to explore, by comparing theoretical predictions and experimental data on $\Delta a_\mu$.

In the subsequent sections, we present the results for three benchmark 
points [{\bf BP}'s], as summarized in Tab.~\ref{tab:1}.
We are interested in the region $M_{Z'} \lesssim 50~{\rm MeV}$ where $Z'$ is sensitive to $\Delta a_\mu$ and  DM annihilation is enhanced due to the longitudinal polarization of the $Z'$.
The region $M_{Z'} \lesssim 10$ MeV is excluded by $\Delta N_{\rm eff}$ measurements.
[{\bf BPI}] and [{\bf BPII}] are chosen to be just below the current 
direct upper bounds on the $U(1)_X$ gauge coupling $g_X$, for which $\Delta a_\mu \simeq 10 \times 10^{-10}$.
[{\bf BPIII}] is chosen to be consistent with the null results from the BMW and the CMD-3, 
while the $Z'$ is light enough to contribute to $\Delta N_{\rm eff}$, thus relaxing the Hubble tension.
Therefore [{\bf BPIII}] will also serve a benchmark if future measurements of the muon 
$(g-2)$ align with the SM predictions.  It is noteworthy that the Hubble tension 
can potentially be alleviated in the case of [{\bf BPI}] and [{\bf BPIII}] 
with the assistance of a light $Z'$ contributing to a certain amount of dark 
radiation \cite{Kamada:2015era, Escudero:2019gzq}.

\begin{table}[htp]
\caption{Three benchmark points we consider in this paper. We also summarize the values of $\Delta a_\mu$ and the 
possibilities for relaxing the Hubble tension with $\Delta N_{\rm eff} \sim 0.2$ for the benchmark points 
{\bf BPI} and {\bf BPIII}. 
}
\label{tab:1}
\begin{center}
\begin{tabular}{|c||c|c|c|}
\hline
        &   $(M_{Z'}, g_X)$ & $ \Delta a_\mu $ & Hubble tension
 \\
 \hline
{\bf BPI} & $(11.5 {\rm MeV}, 4\times 10^{-4})$ &   $15.3 \times 10^{-10}$  &  Yes
\\
{\bf BPII} & $(50 {\rm MeV}, 5\times 10^{-4} )$ &  $12.5 \times 10^{-10}$  & No
\\
{\bf BPIII} & $ (10 {\rm MeV}, 10^{-5} )$ &  $9.9 \times 10^{-13} (\simeq 0)$   & Yes
\\
\hline
\end{tabular}
\end{center}
\end{table}%

\section{The Scalar DM ($X$)}

\subsection{A Generic Case: $Q_X / Q_\Phi \neq  \pm 1, \pm 1/2, \pm 1/3, etc.$} 
Let's first consider complex scalar DM with a generic $Q_X / Q_\Phi$. The Lagrangian governing the gauge invariant and renormalizable interactions of the scalar DM is given by:
\begin{equation}
\mathcal{L}_{\rm DM} = 
|  D_\mu X |^2 -m^2_X |X|^2 - \lambda_{HX} |X|^2 \left( |H|^2 - \frac{v^2_H}{2}  \right) - \lambda_{\Phi X} |X|^2  \left( | \Phi |^2  - \frac{v^2_\Phi}{2}   \right),\label{eq:S_generic}
\end{equation}
where $D_\mu X=(\partial_\mu + i g_X Q_X Z'_\mu) X$ is the covariant derivative. In this formulation, we assume $Q_X = 1$, and $Q_\Phi$ is chosen 
in a way to prevent the presence of gauge-invariant operators up to dimension-5 that would make $X$ decay, 
ensuring the stability of the DM particle \cite{Baek:2013qwa, Ko:2022kvl}. 
This scenario is referred to as the
``generic" case~\footnote{Cases with $Q_\Phi = 2$ or $3$ fall into special categories, as they involve additional operators that contribute to DM stability. Further details are discussed below.}. 

\begin{figure}
\centering
\includegraphics[width=0.45\linewidth]{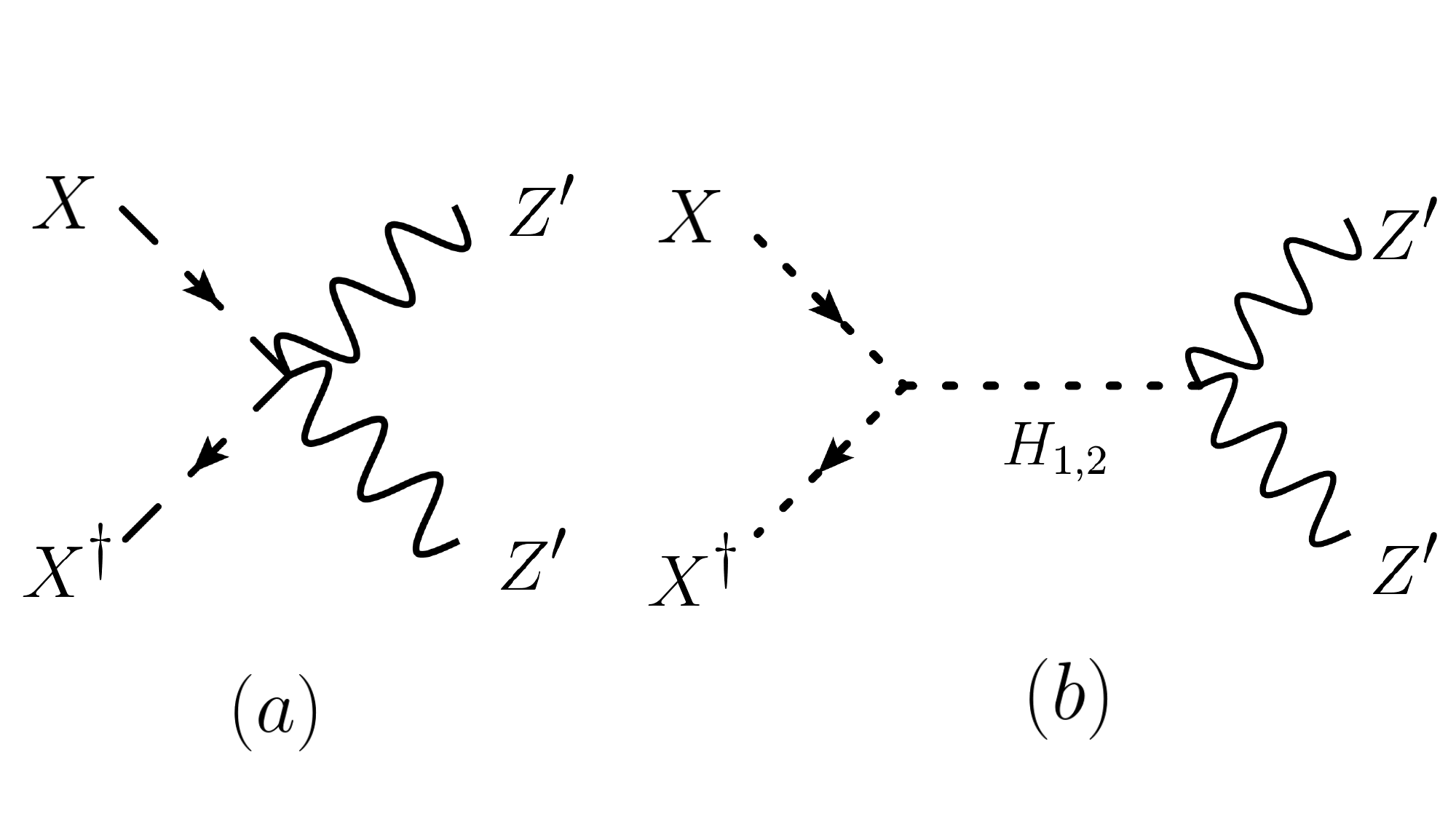}
\includegraphics[width=0.45\linewidth]{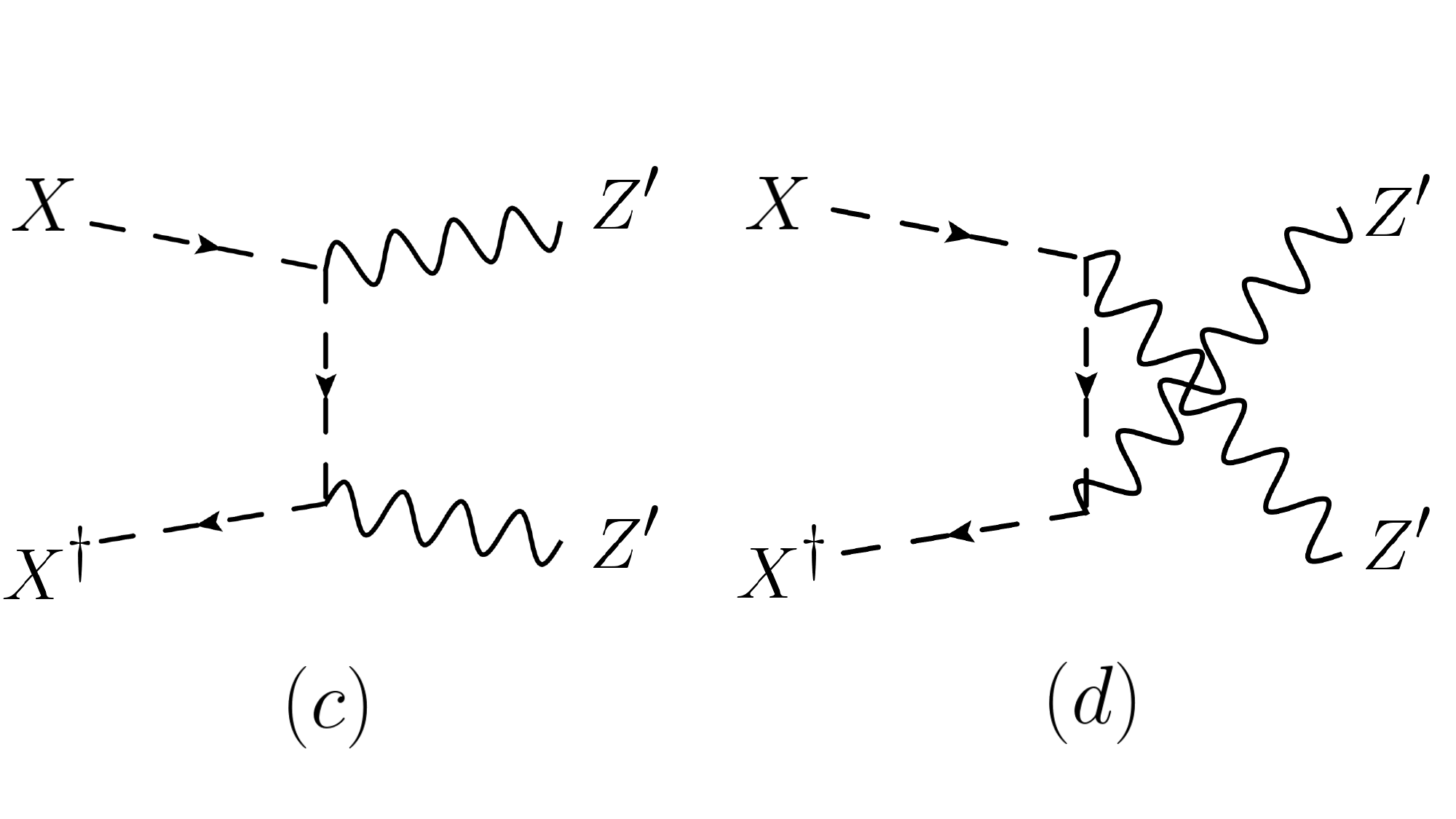}
\hspace{0.5cm}
\includegraphics[width=0.45\linewidth]{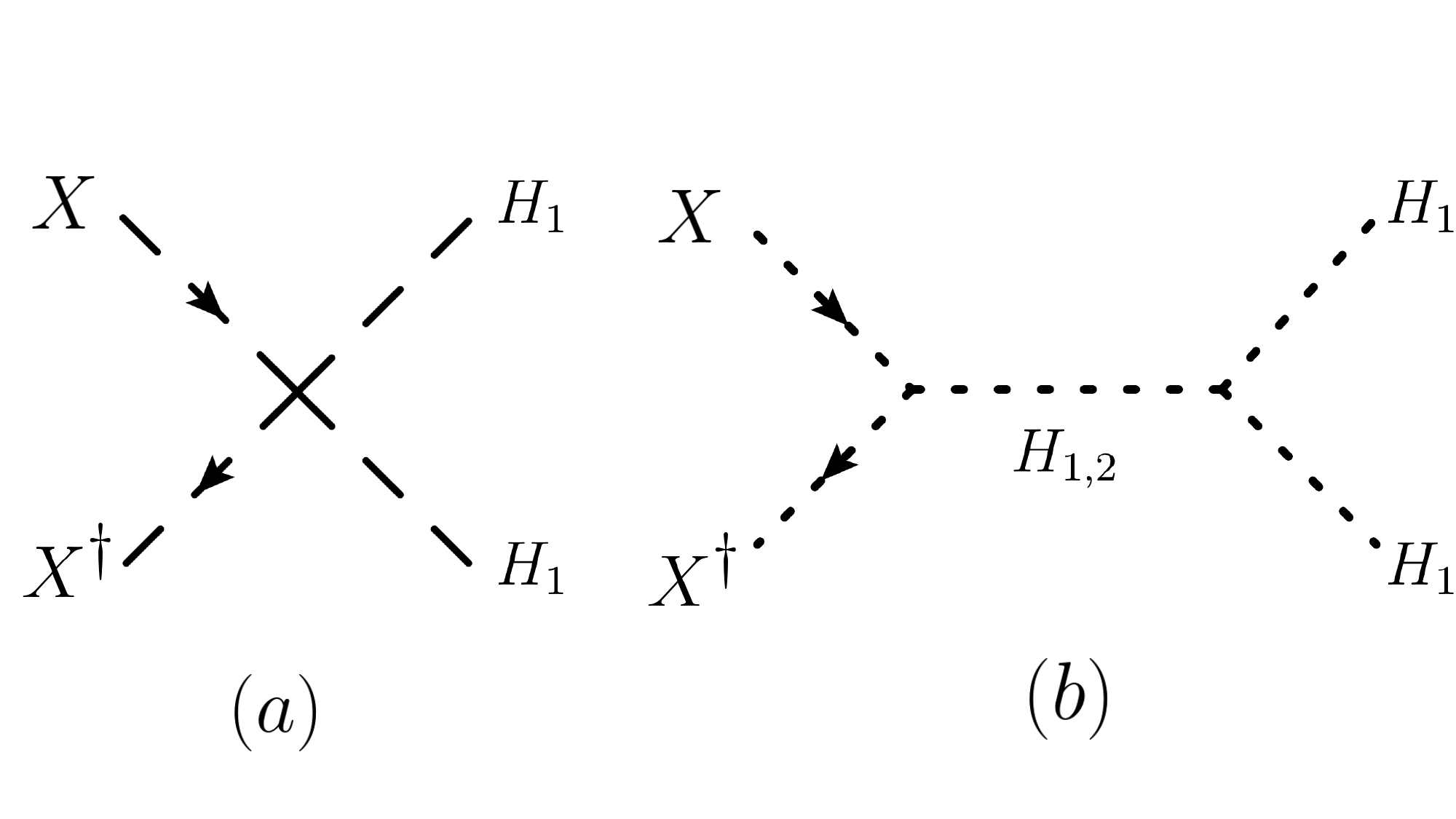}
\includegraphics[width=0.45\linewidth]{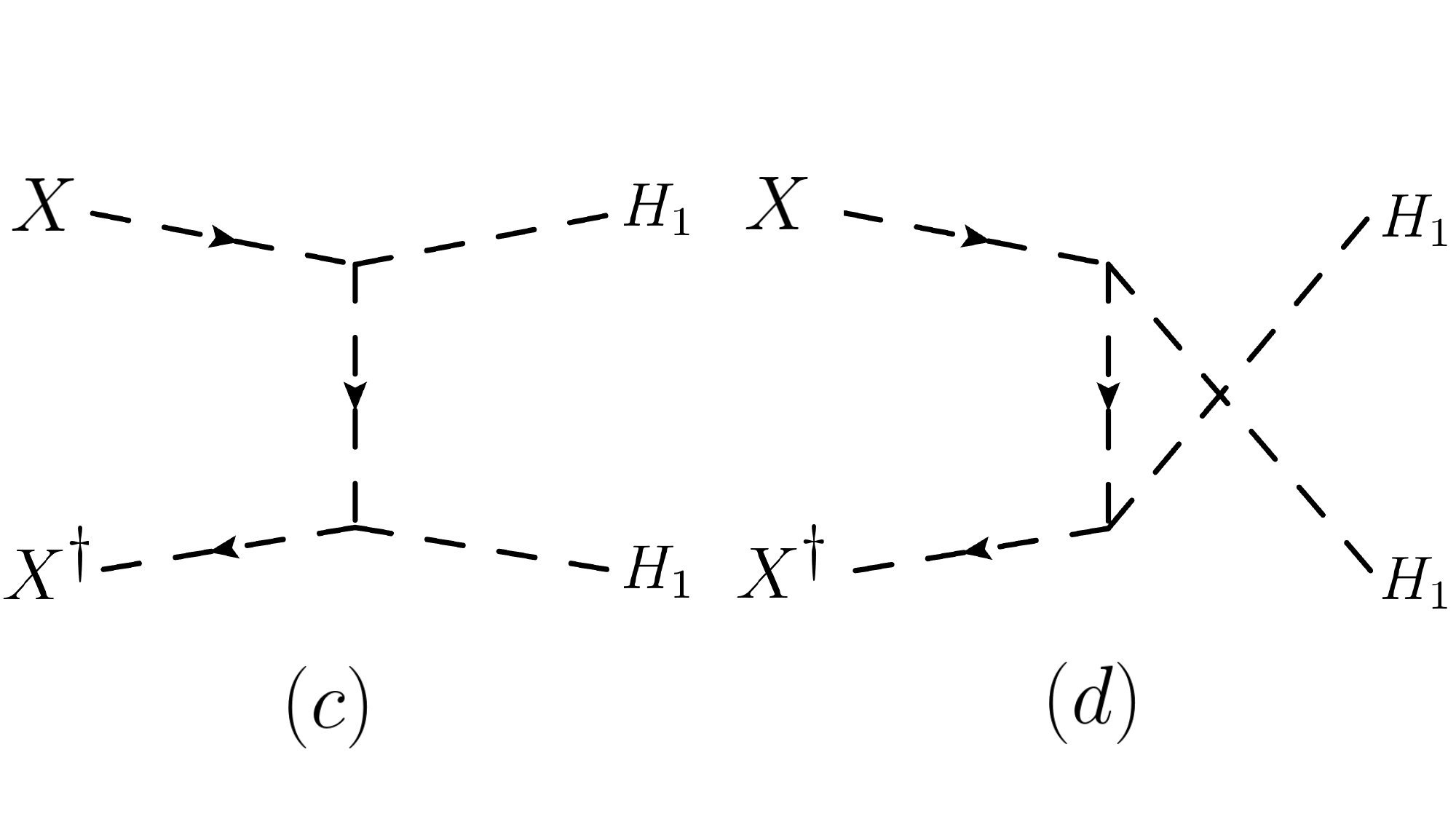}
\hspace{0.5cm}
\caption{ The Feynman diagrams for the  annihilations of complex scalar DMs into a pair of $Z'$ bosons ({\it top})  and $H_1$ bosons ({\it bottom}).  \label{FeynmanSDM} }
\end{figure}

In Fig.~\ref{FeynmanSDM}, we illustrate the Feynman diagrams for the  annihilations of complex scalar DMs into a pair 
of $Z'$ bosons ({\it top})  and $H_1$ bosons ({\it bottom}). They are crucial in achieving the correct DM relic abundance. 
Given the small ${\rm U}(1)_X$ gauge coupling $g_X$, the diagrams $X X^\dagger \rightarrow Z'Z'$ without the $H_1$ contributions are not efficient enough to deplete DM particles. 
This limitation has been observed in prior studies.

The significant enhancement of annihilation cross-sections can occur either by producing longitudinally polarized $Z'$  in the $H_1$-mediated $s$-channel diagram~\cite{Baek:2015fea} [Fig.~\ref{FeynmanSDM} ({\it top}) $(b)$], or by assuming a large value for $\lambda_{\Phi X}$ [Fig.~\ref{FeynmanSDM} ({\it bottom})]. 
For the $XX^\dagger \to Z'Z'$ it is given by
\begin{eqnarray}
\sigma v_{\rm rel} (XX^\dagger \to Z'Z')  &=& \frac{1}{32\pi s} \overline{ \vert \mathcal{M} \vert^2} \left(1 -\frac{4M^2_{Z'} }{s} \right)^{1/2}.\label{eq:XX2ZpZp1}
\end{eqnarray}
The dominant contribution from the $Z'_L Z'_L$ final state gives 
\begin{eqnarray}
\overline{ \vert \mathcal{M} \vert^2} &\simeq & 2 \lambda^2_{\Phi X}  s^2  \left\vert  \frac{1}{s- M^2_{H_1}+i M_{H_1}\Gamma_{H_1} }\right\vert^2.
\end{eqnarray}
In the small $\alpha$ limit, the annihilation cross section is approximated as
\begin{align}
\sigma v_{\rm rel} (XX^\dagger \to Z'Z')  &\approx \left(\lambda^2_{\Phi X} \over 4\pi  \right)  
\frac{ M_X^2 }{\left( 4M^2_X -M^2_{H_1} \right)^2 +M^2_{H_1}\Gamma^2_{H_1}  }. \label{eq:XX2ZpZp}
\end{align}
	
The annihilation cross section of $XX^\dagger \to H_1 H_1$ is given by
\begin{align}
\sigma v_{\rm rel} (XX^\dagger \to H_1 H_1)  &\simeq  \frac{1}{32\pi s}   \left( \lambda_{\Phi X} c^2_\alpha + \lambda_{HX} s^2_\alpha \right)^2 \sqrt{ 1-\frac{4M^2_{H_1}}{s} }. 
\label{eq:XX2H1H1}
\end{align}
Note that Eq.~\eqref{eq:XX2ZpZp} and \eqref{eq:XX2H1H1} become comparable to each other when $\alpha \ll {\cal O}(1)$ and $M_{H_1} \ll M_X$.
This is achieved through the large enhancement, $(M_X/M_{Z'})^4$, in the $H_1$-exchanged $s$-channel diagram of $X X^\dagger \to Z'_L Z'_L$ (See Fig.~\ref{FeynmanSDM} (b)). In principle this enhanced $X X^\dagger \to Z'_L Z'_L$ mode could function even without the dark Higgs. To explore this possibility let us decouple $H_1$ by taking $M_{H_1} \to \infty$. In this limit only diagrams $(a)$ and $(d)$ in Fig.~\ref{FeynmanSDM} are relevant. In this case the dark gauge bosons interact with the DM, however, through the dark gauge interaction, making the annihilation cross section proportional to $g_X^4$, which is very small, ${\cal O}(10^{-13})$, and thus highly suppressed. Therefore, it is essential to introduce the dark Higgs  to allow the 
electroweak-scale dark matter.
	
The thermal WIMP DM can be probed through direct detection experiments, which constrain 
the spin-independent (SI) DM-nucleon scattering cross section. 
The contribution of the light dark Higgs boson $H_1$ to the SI cross section is as follows:
\begin{eqnarray}
\sigma_{\rm SI} &=& \frac{\mu^2_N}{4\pi} \left( \frac{M_N}{M_X} \right)^2 \frac{c^4_\alpha}{M^4_{H_1}} f^2_N \left[ \lambda_{\Phi X} \frac{v_\Phi}{v_H} t_\alpha \left(1-\frac{M^2_{H_1}}{M^2_{H_2}}\right) -\lambda_{HX} \left( t^2_\alpha +\frac{M^2_{H_1}}{M^2_{H_2}}\right) \right]^{2} ,
\end{eqnarray}
where $t_\alpha=\tan\alpha$, $\mu_N$ represents the reduced mass of the dark matter and nucleon system, $M_N$ is the nucleon mass, and $f_N = 0.327$ is a nuclear form factor \cite{Young:2009zb, Crivellin:2013ipa}. The cross section is influenced by several parameters, including the dark Higgs mixing angle $\alpha$, the masses of the  Higgs bosons $M_{H_1}$ and $M_{H_2}$, and the couplings $\lambda_{\Phi X}$ and $\lambda_{HX}$. The most stringent constraints on this cross section come from experiments such as CRESST \cite{CRESST:2019jnq}, DarkSide-50 \cite{DarkSide:2018kuk}, XENON1T \cite{XENON:2018voc,XENON:2019gfn} and XENONnT \cite{XENONCollaboration:2023orw}.
In our study these experimental constraints can be mitigated by choosing small values for $\sin\alpha$ and setting $\lambda_{HX}$ to zero. This allows for a wider parameter space that satisfies both the relic density requirements and the direct detection bounds.

\begin{figure}[!t]
\centering
\includegraphics[width=0.45\linewidth]{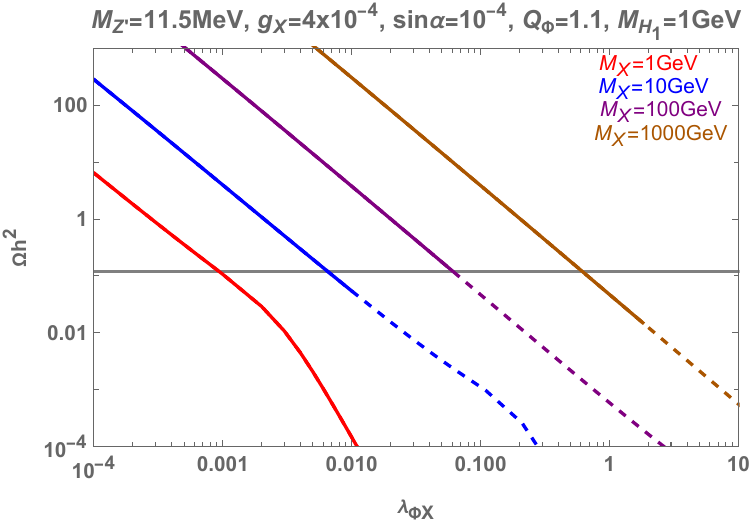}
\includegraphics[width=0.45\linewidth]{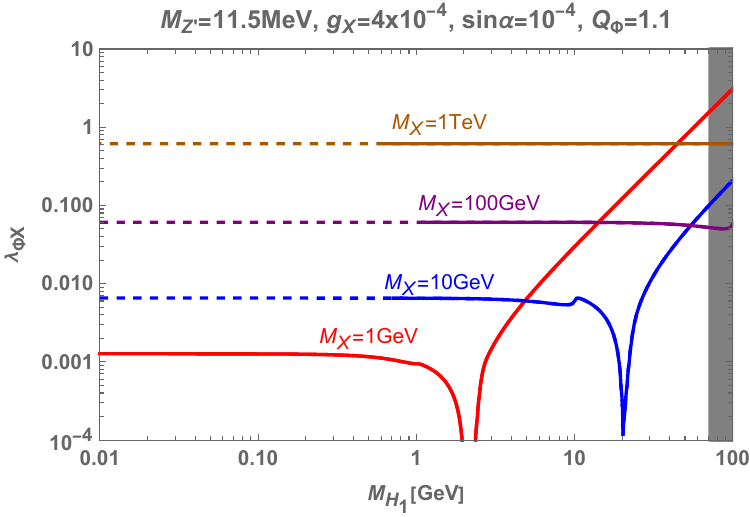}
\caption{Plots in the generic scalar DM model. DM relic abundance $\Omega_{\rm DM} h^2$ as a function of $\lambda_{\Phi X}$ ({\it Left}) and contours for $\Omega_{\rm DM} h^2=0.12$ in the $(M_{H_1}, \lambda_{\Phi X})$-plane ({\it Right}) for [{\bf BPI}]. We take $Q_\Phi=1.1$, $M_{H_1} = 1$ GeV, and $\sin\alpha = 10^{-4}$ for the plots. The curves represent various DM masses: $M_X=1$ (red), $10$ (blue), $100$ (purple), and $1000$ (brown) GeV. The solid (dashed) region is allowed (excluded) by DM direct detection experiments.
	}\label{sclarDMFigBP1}
\end{figure}

In Fig.~\ref{sclarDMFigBP1}  we depict  the impact of the $\lambda_{\Phi X}$ parameter and longitudinally polarized $Z'$ on 
the DM relic density ($\Omega_{\rm DM} h^2$) in the [{\bf BPI}] scenario.
The chosen parameters for these plots are $Q_\Phi=1.1$, $M_{H_1} = 1$ GeV, and $\sin\alpha = 10^{-4}$. 
The curves represent different DM masses: $M_X=1$ (red), $10$ (blue), $100$ (purple), and $1000$ (brown) GeV.
The solid (dashed) region is allowed (excluded) by DM direct detection experiments.

In the left panel, the DM relic abundance, denoted as $\Omega_{\rm DM} h^2$, is plotted as a function of $\lambda_{\Phi X}$. 
The observed relic density, represented by a horizontal gray line, is $\Omega_{\rm DM} h^2 = 0.12$. 
For a given $M_{Z'}$, the annihilation cross sections $X X^\dagger \to Z' Z'$ (or $H_1 H_1$) scale as $\lambda_{\Phi X}^2/M_X^2$. 
This relation elucidates the observed trends in the graph.

In the right panel, contours corresponding to $\Omega_{\rm DM} h^2=0.12$ are shown in the $(M_{H_1}, \lambda_{\Phi X})$-plane. 
The primary diagram contributing to the total cross-section for the annihilation process $XX^\dagger \to Z'Z'$ is illustrated in Fig.~\ref{FeynmanSDM} (Top)(b). 
In the scenario [{\bf BPI}], the relic density is predominantly governed by the processes $X X^\dagger \rightarrow H_1 H_1$ and $Z'Z'$ when $M_{H_1} < 2 M_X$. 
For  $M_{H_1} \geq 2 M_X$, the annihilation into $Z'Z'$ becomes the main contributor to the relic density 
as the channel $X X^\dagger \rightarrow H_1 H_1$ becomes kinematically inaccessible. 
The sensitivity of the dominant annihilation cross-section terms to $M_{H_1}$ changes notably across the resonance threshold; below this threshold, the change in $M_{H_1}$ has minimal impact because $\langle \sigma v_{\rm rel}\rangle_{Z' Z' (H_1 H_1)} \propto (\lambda_{\Phi X}/M_X)^2$, 
whereas above it, $\langle \sigma v_{\rm rel}\rangle_{Z' Z'}$ is proportional to $(\lambda_{\Phi X} M_X/M^2_{H_1})^2$. 
This  explains the distinct behaviors observed in the plot for regions below and above the $H_1$ resonance.

From these analyses, it emerges that DM masses significantly deviating from the relation~(\ref{eq:res}) can still achieve the correct thermal relic density if the dark Higgs boson contributions are included, 
highlighting a critical insight from our study.

\begin{figure}[!h]
\centering
\includegraphics[width=0.45\linewidth]{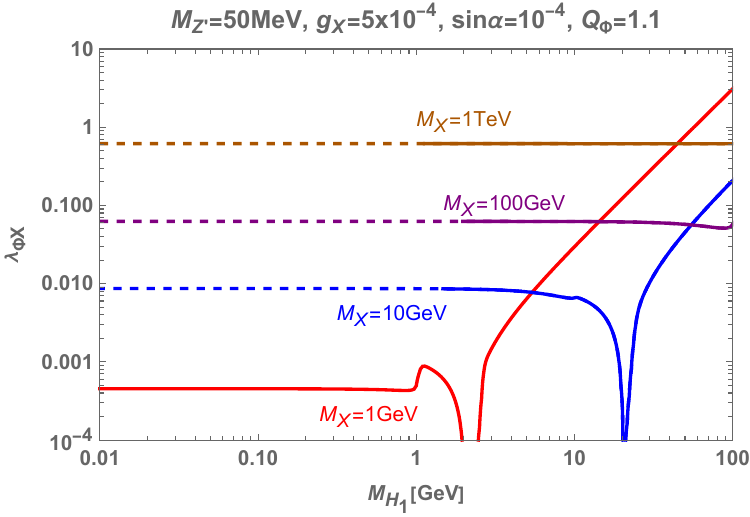}
\includegraphics[width=0.45\linewidth]{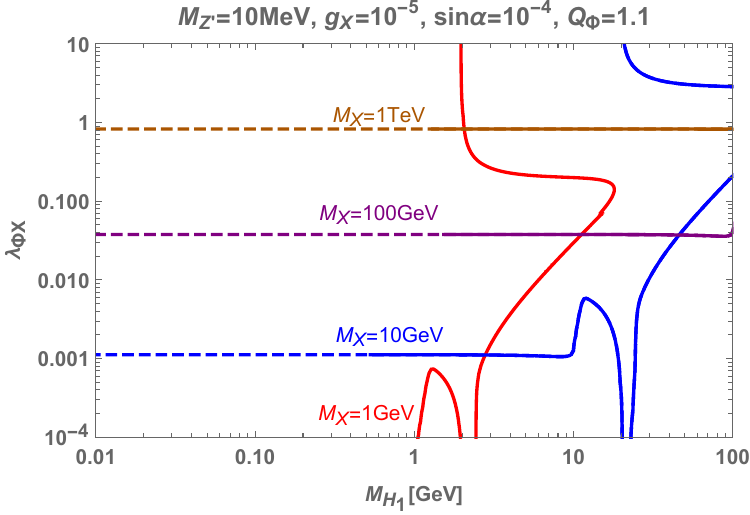}
\caption{ The plots corresponding to the right side of Fig.~\ref{sclarDMFigBP1} for [{\bf BPII}] (left) and [{\bf BPIII}] (right). }		
\label{sclarDMFigBP2}.
\end{figure}

In Fig. \ref{sclarDMFigBP2}, we present plots analogous to those in Fig.~\ref{sclarDMFigBP1} 
for the benchmark points [{\bf BPII}] and [{\bf BPIII}].
The plots reveal that scalar DM can be a thermal WIMP over a broad mass range. 
This behavior is attributed to contributions from the dark Higgs $\phi \simeq H_1$, 
introducing novel channels such as $Z' Z'$ and a distinct annihilation channel into $H_1 H_1$.
Notably, the scalar DM's thermal properties extend beyond the region in Eq.~(\ref{eq:res}).
For $M_X \lesssim 10$~GeV in the case of [{\bf BPIII}],  there is unique behavior after the dark Higgs resonance. 
This behavior is originated from the dark Higgs boson decay width in the denominator of $XX^\dagger \to Z'Z'$ annihilation cross section.
The dark Higgs decays dominantly into a pair of DMs, and its decay width is proportional to 
$\lambda_{\Phi X}^2 v^2_\Phi / M_{H_1}$.  In the denominator, the term $\Gamma^2_{H_1} M^2_{H_1}$ can be dominant over the mass of the dark Higgs when $\lambda_{\Phi X}$ is large. Therefore, we can obtain two 
distinct solutions for a given $M_{H_1}$, explaining the pattern.	

Before closing this subsection, we wish to emphasize that the strict relation (\ref{eq:res}) can 
be largely circumvented by incorporating the dark Higgs sector.  This can be achieved through
 $X X^\dagger \to H_1 H_1$ and/or $X X^\dagger \to Z' Z'$, where $Z'$'s are longitudinally polarized. Both processes are primarily controlled by $\lambda_{\Phi X}$ and $M_{H_1}$. 
 These mechanisms operate independently of the $Z'$ explanation for $\Delta a_\mu$ because $g_X$ is not directly involved in either of the annihilation processes.


\subsection{Local $Z_2$ scalar DM: $(Q_X , Q_\Phi ) = (1,2)$}
Let us consider a special case with charge assignments $Q_\Phi = 2$ and $Q_X = 1$. In this scenario, the DM Lagrangian is augmented by an additional gauge-invariant operator at the renormalizable level, expressed as:
\begin{equation}
\Delta \mathcal{L}_{\rm DM} = - \mu ( X^2 \Phi^\dagger + \text{H.c.})
\end{equation}
This term supplements the generic dark matter Lagrangian specified in Eq. (\ref{eq:S_generic}). Consequently, the ${\rm U}(1)_X$ symmetry is spontaneously broken down to its $Z_2$ subgroup ($X \rightarrow -X$),  {\it \'{a} la} 
Krauss-Wilczek mechanism \cite{Krauss:1988zc}, after $\Phi$ gets nonzero VEV.
This phenomenon has been previously explored in the literature for explaining anomalies such as the Galactic Center $\gamma$-ray excess~\cite{Baek:2014kna, Okada:2019sbb} 
and the XENON1T excess~\cite{Baek:2020owl}.

Upon symmetry breaking induced by a nonzero $v_\Phi$, the $\mu$-term is rewritten as:
\begin{equation}
\mu (X^2\Phi^\dagger + \text{H.c.}) = \frac{1}{\sqrt{2}} \mu v_\Phi (X^2_R - X^2_I) \left( 1+ \frac{\phi}{v_\Phi} \right),
\end{equation}
where $X=(X_R + i X_I)/\sqrt{2}$. This interaction leads to a mass splitting between the real ($X_R$) and imaginary ($X_I$) components of $X$:
\begin{eqnarray}
M^2_{R} &=& M^2_X +\sqrt{2} \mu v_\Phi, \nonumber\\
M^2_{I} &=& M^2_X -\sqrt{2} \mu v_\Phi.
\end{eqnarray}
Assuming $\mu > 0$, the lighter state $X_I$ becomes the DM candidate. 
The mass splitting is quantified by the dimensionless parameter $\Delta \equiv ( M_R-M_I)/M_I$. Notably, the interaction between the dark photon and 
DM is off-diagonal (or inelastic) and described by
\begin{eqnarray}
\mathcal{L} \supset g_X Z'_\mu ( X_R \partial^\mu X_I - X_I \partial^\mu X_R ).
\end{eqnarray}
This introduces intriguing aspects to the DM phenomenology, particularly regarding the mass splitting's implications on DM stability and detection prospects.

\begin{figure}
\centering
\includegraphics[width=0.45\linewidth]{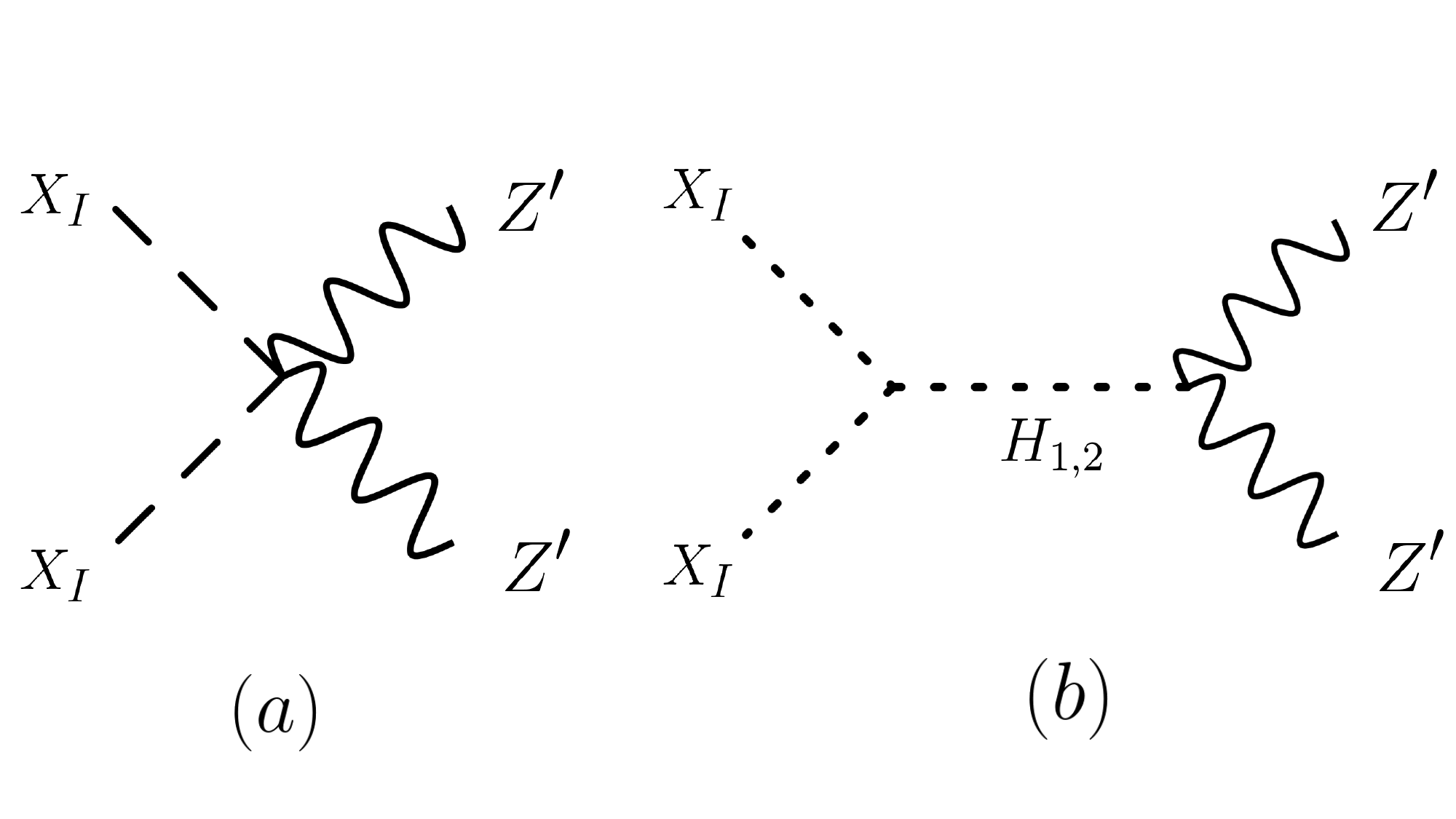}
\includegraphics[width=0.45\linewidth]{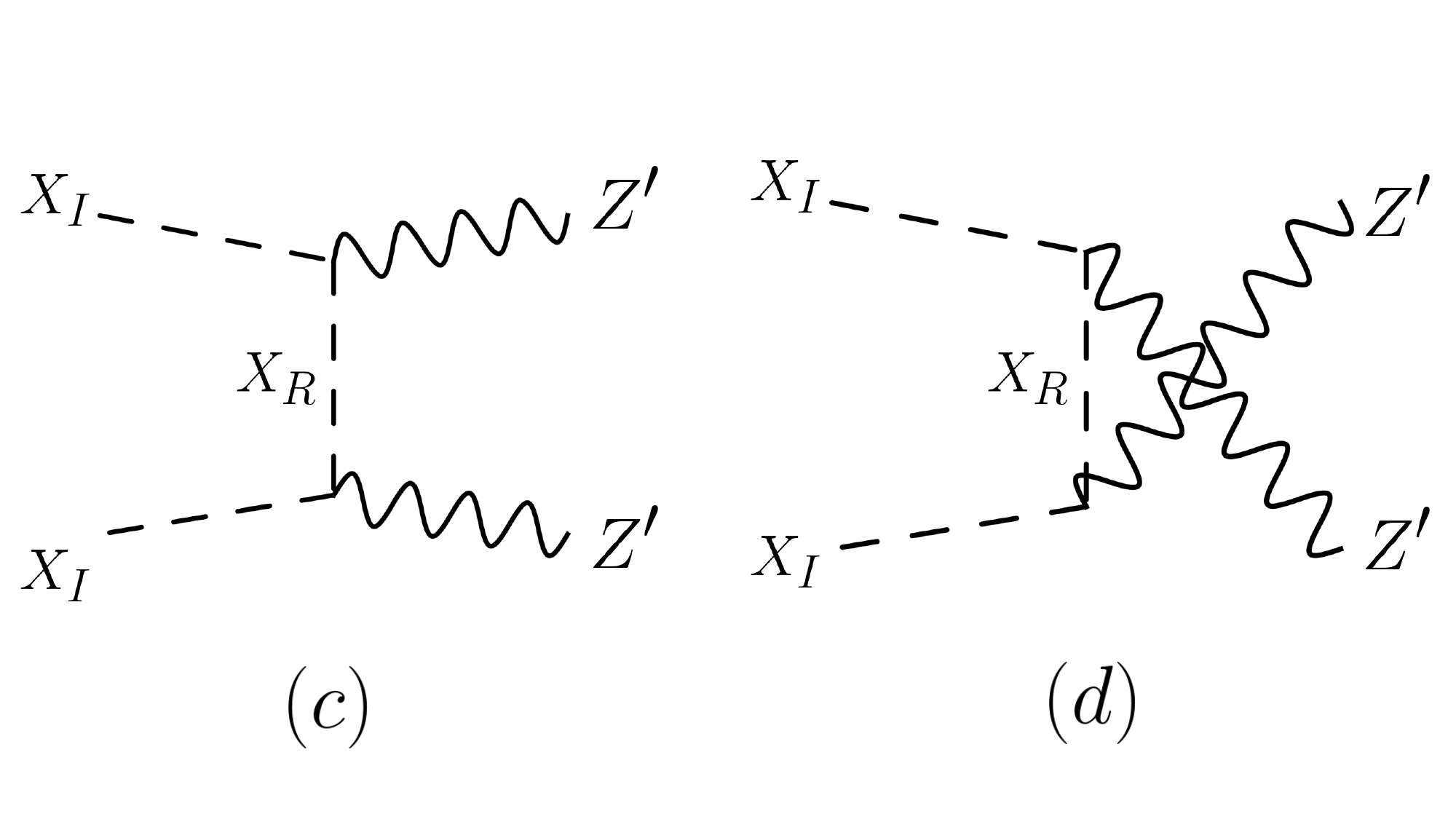}
\hspace{0.5cm}
\includegraphics[width=0.45\linewidth]{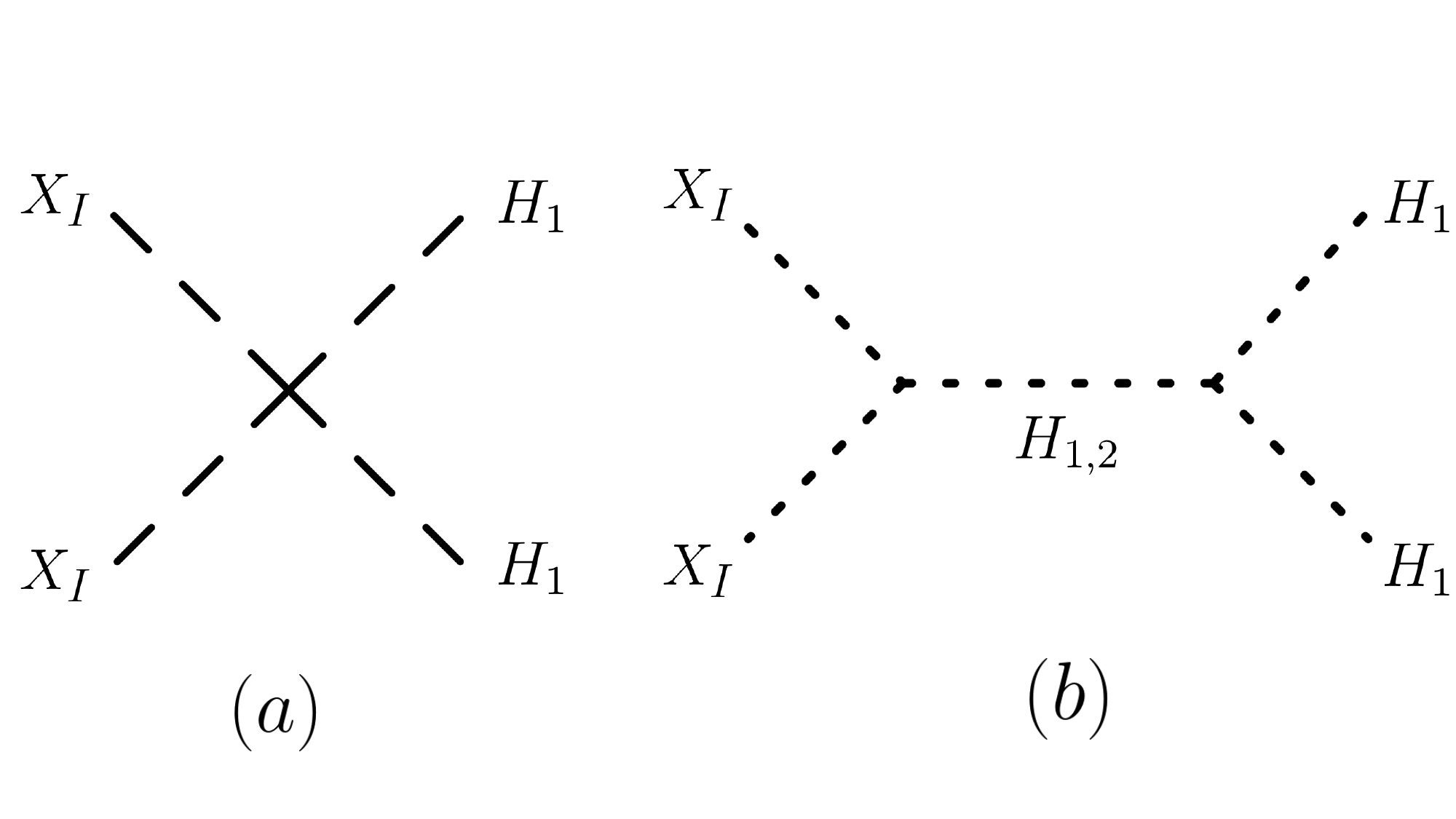}
\includegraphics[width=0.45\linewidth]{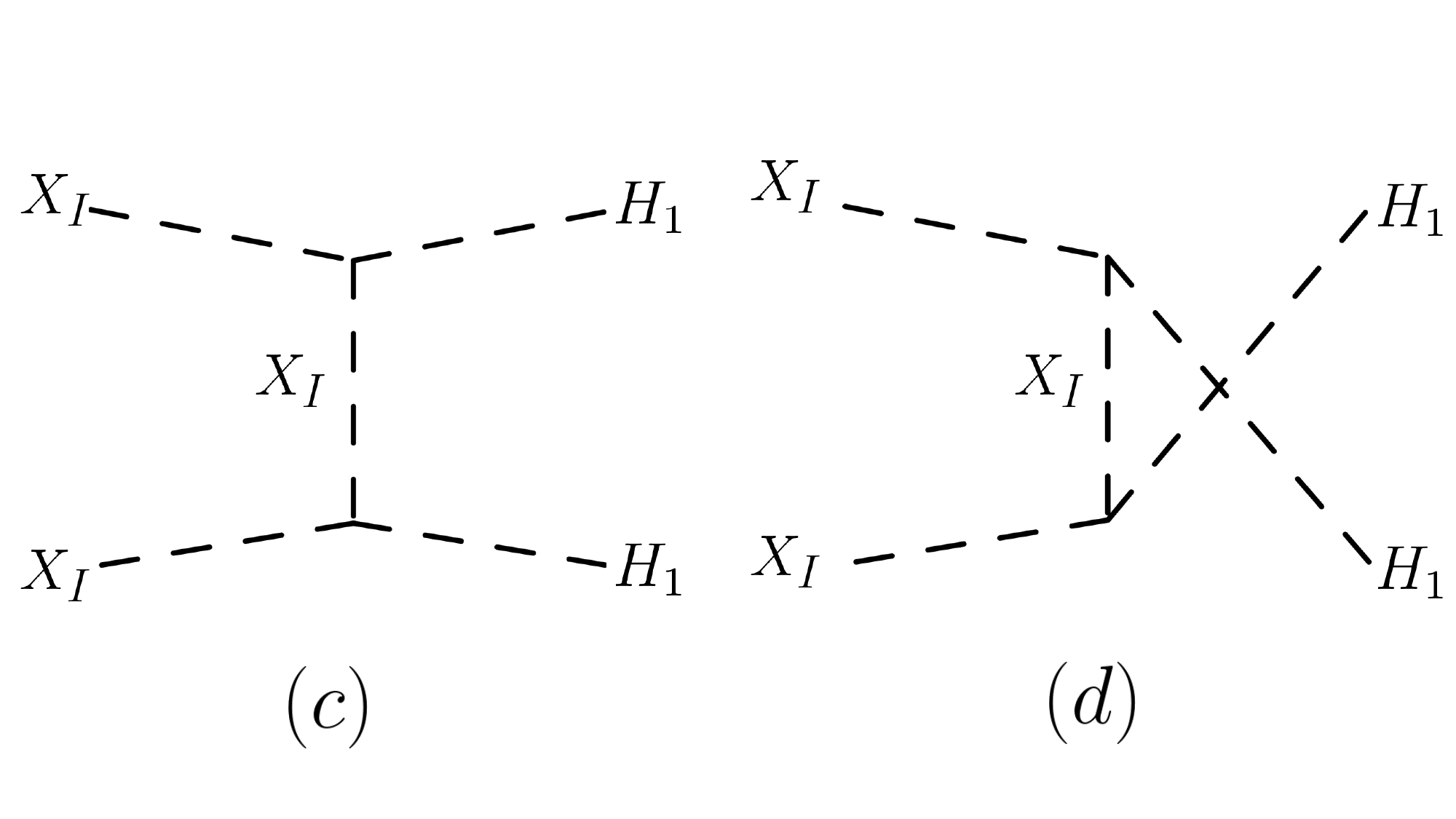}
\hspace{0.5cm}
\caption{  Feynman diagrams for the local $Z_2$ scalar DM model annihilating to a pair of $Z'$ bosons ({\it Top}) and  to a pair of $H_1$ bosons ({\it Bottom}) \label{FeynmanZ2SDM}.} 
\end{figure}

For the benchmark point {\bf [BPI]},  
there are two dominant DM annihilation channels: $X_I X_I \to Z' Z', H_1 H_1$, 
similar to the generic case discussed in the previous subsection.
The corresponding Feynman diagrams are  depicted in Fig.~\ref{FeynmanZ2SDM}. 

The cross-section for annihilation into a pair of $Z'$ bosons is expressed as
\begin{equation}
\sigma v_{\rm rel} (X_I X_I \to Z'Z') = \frac{1}{32\pi s} \overline{|\mathcal{M}|^2} \left( 1 - \frac{4M^2_{Z'}}{s}\right)^{1/2},
\end{equation}
where the averaged squared matrix element $\overline{|\mathcal{M}|^2}$ is approximated by
\begin{equation}
\overline{|\mathcal{M}|^2} \simeq \frac{s^2}{v^2_\Phi} \left| \frac{\lambda_1 c_\alpha }{s-M^2_{H_1} +i\Gamma_{H_1} M_{H_1}} + \frac{\lambda_2 s_\alpha }{s-M^2_{H_2} +i\Gamma_{H_2} M_{H_2}} \right|^2,
\end{equation}
with $\lambda_1 = (\lambda_{\Phi X} v_\Phi -\sqrt{2} \mu) c_\alpha - \lambda_{HX} v_H s_\alpha$ and $\lambda_2 = (\lambda_{\Phi X} v_\Phi -\sqrt{2} \mu) s_\alpha + \lambda_{HX} v_H c_\alpha$.
	
It is noted that the contribution from the longitudinal mode of $Z'$ pairs, $Z'_L$, dominates here again. 
Disregarding the small mass of $Z'$ in the final states, the annihilation cross-section can be described as
\begin{equation}
\sigma v_{\rm rel} (X_I X_I \to Z'Z')  \approx \frac{(\lambda_{\Phi X}-\sqrt{2}\mu/v_\Phi)^2}{8 \pi } \frac{M^2_I}{( 4M_I^2 -M_{H_1}^2 )^2 +M^2_{H_1}\Gamma^2_{H_1}}.
\end{equation}
In the $\mu \to 0$ limit, we recover the generic case Eq.~\eqref{eq:XX2ZpZp} by adding the same contribution from $X_R X_R \to Z' Z'$.  

Moreover, if kinematically allowed, dark matter particles can also annihilate into a pair of dark Higgs bosons, $H_1 H_1$, with the cross-section given by
\begin{equation}
\sigma v_{\rm rel} (X_I X_I \to H_1 H_1)  \approx \frac{1}{32\pi s} \left( \lambda_{\Phi X} c^2_\alpha + \lambda_{HX} s^2_\alpha \right)^2 \sqrt{ 1-\frac{4M^2_{H_1}}{s} }.
\end{equation}

\begin{figure}[!h]
\centering
\includegraphics[width=0.45\linewidth]{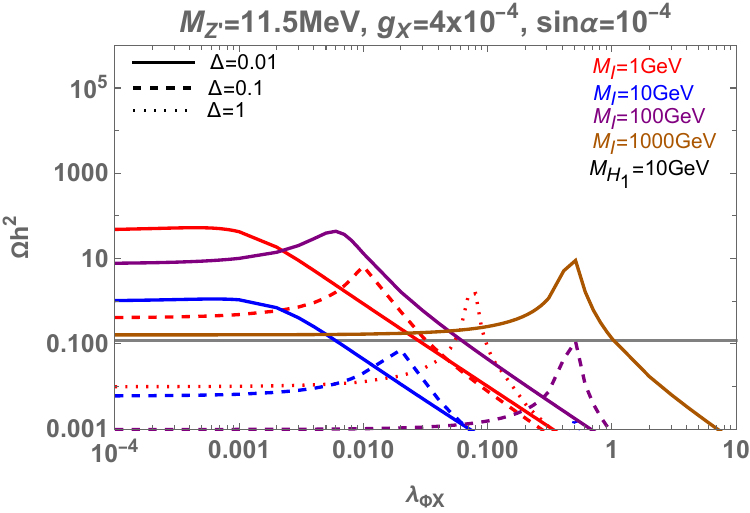}
\includegraphics[width=0.45\linewidth]{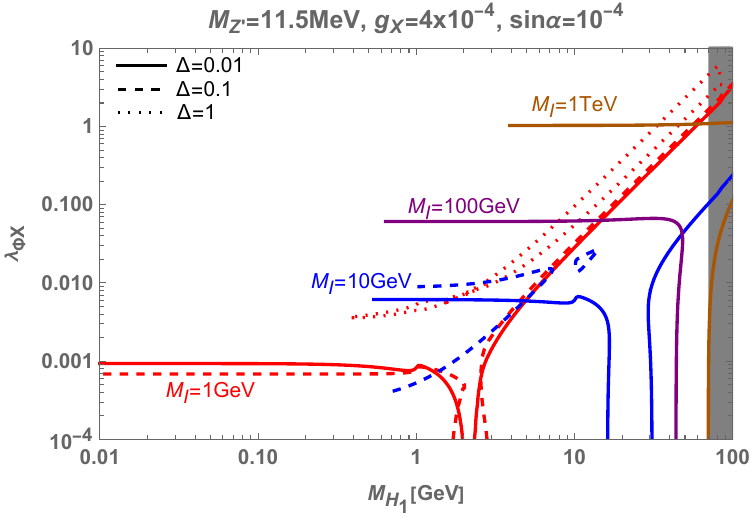}	
\caption{Plots in the local $Z_2$ scalar DM model. 	{\it Left}: Relic abundance of local $Z_2$ scalar DM as a function of $\lambda_{\Phi X}$ for [{\bf BPI]} and different values of mass splitting parameter $\Delta$. We take $\lambda_{HX}=0$, $M_{H_1}=10$~GeV, and $s_\alpha=10^{-4}$. All the curves satisfy the DM direct detection bound.  {\it Right}: The contours of $\Omega_{\rm DM} h^2 =0.12$ in the  		$( M_{H_1} , \lambda_{\Phi X} )$-plane for different values of $\Delta$. The gray area is excluded by the perturbativity condition. 	\label{fig:Z2ScalarDM}} 

\end{figure}


In the left panel of Fig. \ref{fig:Z2ScalarDM}, we present the DM relic density as a function of $\lambda_{\Phi X}$ for the benchmark point {\bf [BPI]}. 
Various scenarios are considered with $M_I$ values ranging from 1 GeV to $10^3$ GeV and $\Delta$ values of 1, 0.1, and 0.01. The parameters $M_{H_1}$ and $\sin\alpha$ are fixed at 10 GeV and $10^{-4}$, respectively. The introduction of the new parameter $\mu$ makes predictions for the relic density distinct from the generic case discussed in the preceding section.

For lighter DM, the correct relic density is achieved through the $X_I X_I \to Z' Z'$ channel, while for heavier DM, the $X_I X_I \to H_1 H_1$ channel becomes relevant. 
In the left panel of Fig. \ref{fig:Z2ScalarDM}, an increase in $\Delta$ strengthens the DM coupling with $H_1$,  leading to a reduction in the relic abundance. 
A noticeable bump-shaped feature is observed, attributed to the cancellation between 
$\lambda_{\Phi X} v_\Phi$ and $\mu$ in Eq. (3.13).

In the right panel of Fig. \ref{fig:Z2ScalarDM}, contours of $\Omega_{\rm DM}=0.12$ are illustrated in the $(M_{H_1},\lambda_{\Phi X})$-plane.
The dominance of the $X_I X_I \to H_1 H_1, Z'Z'$ channels is evident below the resonance region, $M_X \sim M_{H_1}/2$. 
Above the resonance, the $X_I X_I \to Z'Z'$ channel takes precedence. 
Additionally, near the resonance region, the contribution from co-annihilation $X_I X_R \to Z' H_1$ influences the relic density, 
although this effect remains subdominant. For smaller $\Delta$ values, such as $\Delta  \lesssim 10^{-2}$, 
the correct relic density for heavier DM up to a few TeV can be obtained through the $X_I X_I (X_R X_R) \to H_1 H_1, Z'Z'$ channels.

\begin{figure}[!h]
\centering
\includegraphics[width=0.45\linewidth]{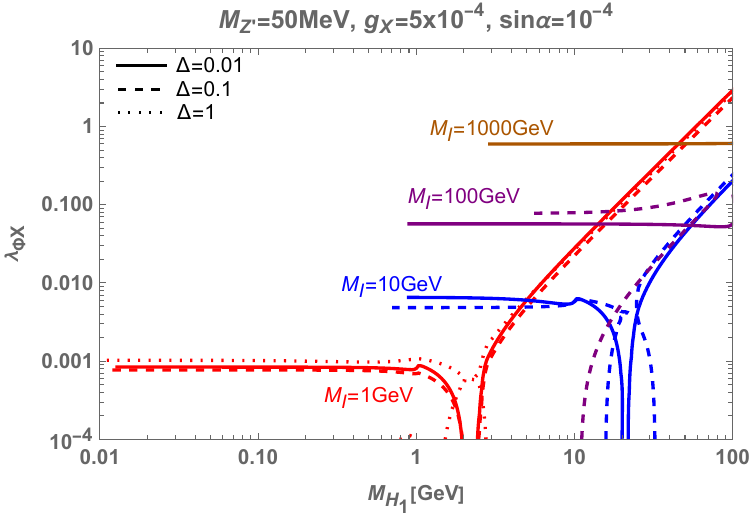}
\includegraphics[width=0.45\linewidth]{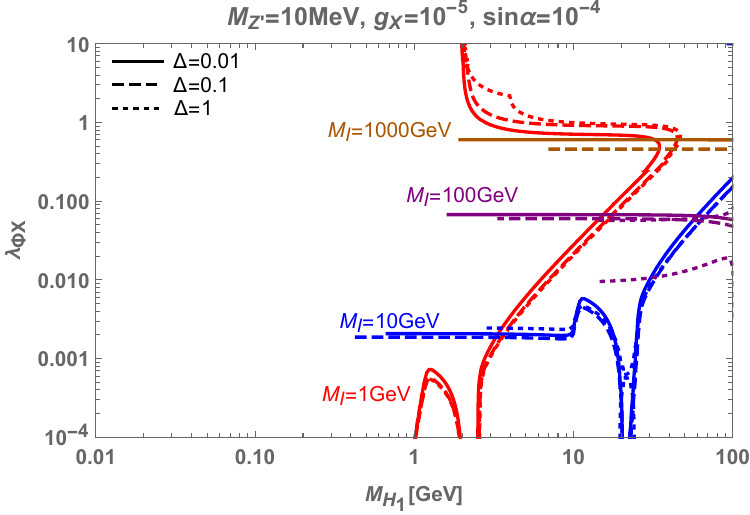}		
\caption{ 
The plot corresponding to Fig.~\ref{fig:Z2ScalarDM} ({\it right}) for the benchmark point [{\bf BPII}] (left) and [{\bf BPIII}] (right). All the lines satisfy the DM direct detection bound. \label{fig:Z2ScalarDM_BP2}
	}
\end{figure}
In Fig. \ref{fig:Z2ScalarDM_BP2}, we show a plot corresponding to Fig.~\ref{fig:Z2ScalarDM} ({\it right}) for the benchmark point [{\bf BPII}]. 
We observe behavior similar to that in Fig.~\ref{sclarDMFigBP2}, where a given $M_{H_1}$ can correspond to two distinct values of $\lambda_{\Phi X}$.

Within the $Z_2$ symmetric scalar dark matter model, dark matter direct detection can proceed through elastic and inelastic scattering processes. Given the parameter space of interest where $M_R -M_I \geq \mathcal{O}(100)$~keV, inelastic scattering does not happen. Therefore, the focus is on elastic scattering mediated by both the dark and the SM Higgs bosons, where the spin-independent elastic dark matter-nucleon scattering cross-section is formulated as
\begin{align}
\sigma_{\rm SI} &= \frac{\mu^2_N}{4\pi} \left( \frac{M_N}{M_I} \right)^2 \frac{c^4_\alpha}{M^4_{H_1}} f^2_N \left[ \left(\lambda_{\Phi X} -\frac{\sqrt{2}\mu}{ v_\Phi}\right) \frac{v_\Phi}{v_H} t_\alpha \left(1-\frac{M^2_{H_1}}{M^2_{H_2}}\right) -\lambda_{HX} \left( t^2_\alpha +\frac{M^2_{H_1}}{M^2_{H_2}}\right) \right]^2.
\end{align}
To circumvent stringent constraints from direct detection experiments, the parameter $\sin\alpha$ is selected to be sufficiently small ($\sim 10^{-4}$), 
while setting $\lambda_{HX}=0$. 
This approach ensures compliance with current experimental bounds while maintaining the viability of the scalar dark matter model under investigation.

\subsection{Local $Z_3$ scalar DM:  $(Q_X , Q_\Phi ) = (1,3)$}
In a specialized scenario for complex scalar dark matter (DM) with charge assignments \(Q_\Phi = 3\) and \(Q_X = 1\), the \({\rm U(1)}_X\) symmetry is reduced to \(Z_3\) through the Krauss-Wilczek mechanism~\cite{Ko:2014loa,Ko:2014nha}. The pertinent Lagrangian for this dark matter model is described by
\begin{align}
\mathcal{L}_{\rm DM} &= D^\mu X^\dagger D_\mu X - m^2_X X^\dagger X - \lambda_{HX} X^\dagger X \left( H^\dagger H - \frac{v^2_H}{2} \right) \nonumber  \\ 
& - \lambda_{\Phi X} X^\dagger X \left( \Phi^\dagger \Phi 
- \frac{v^2_\Phi}{2} \right) + \lambda_3 \left(X^3\Phi^\dagger + \text{H.c.} \right) ,
\end{align}
where the last $\lambda_3$ term is a new gauge invariant operator specific for the charge assignments 
$Q_\Phi = 3 Q_X$.
This model framework introduces unique semi-annihilation processes \(XX\rightarrow X^\dagger H_1, X^\dagger Z'\), supplementing the conventional annihilation pathways \(X X^\dagger \rightarrow Z' (\phi , h) \rightarrow\) SM particles~\cite{Ko:2014nha}. These additional channels allow for a broader mass spectrum for the complex scalar DM \(X\), diverging from the mass correlation (\ref{eq:res}): \(M_X \sim M_{Z'}/2 \) typically inferred in scenarios solely involving \(Z'\) without \(H_1\). 

\begin{figure}[thb]
\centering
\includegraphics[width=0.6\linewidth]{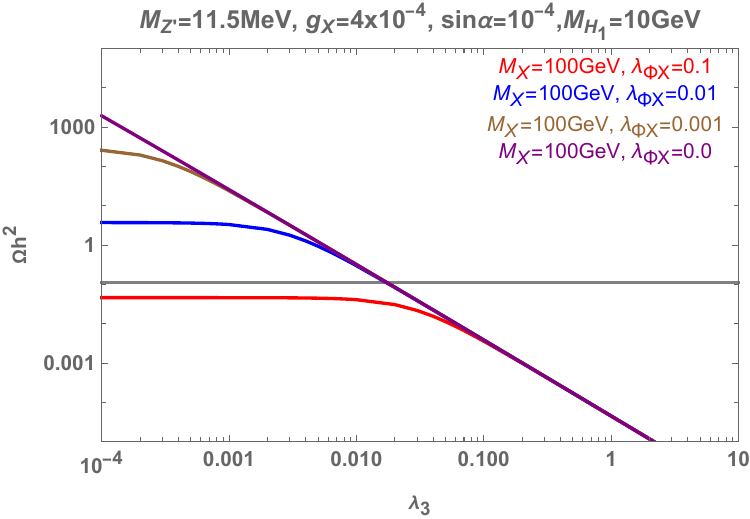}	
\caption{ Relic abundance of $Z_3$ scalar DM for the [{\bf BPI}].   Here we fixed $\lambda_{HX}=0$ for simplicity. \label{fig:z3}}
\end{figure}

In Fig.~\ref{fig:z3}, we show $\Omega_X h^2$ as a function of $\lambda_3$ which controls the strength of semi-annihilation for  different choices of $M_X$ and $\lambda_{\Phi X}$. 
Given the smallness of the gauge coupling $g_X \sim O(10^{-4})$, the semi-annihilation channel $XX \rightarrow X^\dagger Z'$ is less significant than the process $XX \rightarrow X^\dagger H_1$ which is controlled by the parameter  $\lambda_3$.
Once again, we observe that the dark Higgs can modify DM phenomenology  significantly and the allowed mass range for the complex scalar DM $X$ can 
significantly deviate from $M_X \sim M_{Z'}/2$.

\section{The Fermion DM ($\chi$)}

\subsection{Generic Case: $Q_\chi/Q_\Phi \neq  \pm 1/2$}
\label{sec:fDM_generic}
For the generic case of Dirac fermion dark matter (DM), the DM Lagrangian is expressed as
\[
\mathcal{L}_{\rm DM} = \overline{\chi} ( i \slashed{D} - m_{\chi} ) \chi.
\]
In this scenario, direct renormalizable interactions between $\chi$ and $\Phi$ are absent. This situation stands in stark contrast to the scalar DM cases previously discussed. Assuming a gauge coupling $g_X \sim \mathcal{O}(10^{-4})$, the DM pair annihilation cross sections for processes $\chi \bar{\chi} \rightarrow Z' Z'$ and $Z' H_1$ are on the order of $\mathcal{O}(g_X^4/M_{\chi}^2)$. Consequently, these cross sections are too small to yield the correct thermal relic density of DM for $M_\chi \gg M_{Z'}$. The desired relic density can only be achieved in the resonance region, $M_\chi \sim M_{Z'}/2$, via $\chi \bar{\chi}\rightarrow Z' \rightarrow$ SM particles, confirming findings from previous studies.
	
Specifically the thermal-averaged DM annihilation cross section into leptons is given by
\[
\langle \sigma v_{\rm rel} (\chi\bar{\chi} \to Z' \to \ell\bar{\ell}) \rangle \simeq \frac{g^4_X Q^2_\chi}{2\pi } \frac{ \left(2M^2_\chi + M^2_\ell \right)  }{ \left( 4M^2_\chi -M^2_{Z'} \right)^2 +\Gamma^2_{Z'} M^2_{Z'} } \sqrt{ 1- \frac{M^2_\ell}{M^2_\chi} },
\]
and similarly for neutrinos,
\[
\langle \sigma v_{\rm rel} (\chi\bar{\chi} \to Z' \to \nu_\ell \bar{\nu_\ell}) \rangle \simeq \frac{g^4_X Q^2_\chi}{2\pi } \frac{ M^2_\chi  }{ \left( 4M^2_\chi -M^2_{Z'} \right)^2 +\Gamma^2_{Z'} M^2_{Z'} },
\]
where $\ell=\mu,\tau$.
For $g_X \sim \mathcal{O}(10^{-4})$ and $M_{Z'} \sim (10-100)$ MeV, the total annihilation cross section falls short of achieving the correct relic density in the off-resonance region. 
An appropriate DM relic abundance is attainable only in the resonance region with $M_\chi \sim M_{Z'}/2$.

If kinematically allowed, additional DM annihilation channels include:
\begin{align}
\langle \sigma v_{\rm rel} (\chi\bar{\chi} \to Z' Z')  \rangle &\simeq \frac{g^4_X Q^4_\chi}{4\pi}\frac{M^2_\chi - M^2_{Z'}}{\left( M^2_{Z'} - 2M^2_\chi  \right)^2 } \sqrt{1 - \frac{M^2_{Z'}}{M^2_\chi}} \\
\langle \sigma v_{\rm rel} (\chi\bar{\chi} \to Z' H_1)  \rangle &\simeq \frac{c^2_\alpha g^4_X Q^2_\Phi Q^2_\chi }{256\pi } \frac{ \left[16 M_\chi^4 -8 M_\chi^2 (M^2_{H_1} -5 M^2_{Z'})  + \left( M^2_{H_1} - M^2_{Z'} \right)^2 \right]  }{ M_\chi^4 \left[ \left( 4M^2_\chi - M_{Z'}^2  \right)^2 + M^2_{Z'}\Gamma^2_{Z'} \right]} \nonumber\\
&\times \sqrt{M^4_{H_1} + \left( 4M^2_\chi - M^2_{Z'} \right)^2  -2 M^2_{H_1} \left( M^2_{Z'} +4 M^2_\chi \right) }
\end{align}
Nevertheless, achieving the correct relic density of DM is constrained primarily to the vicinity of the resonance region because of the too-small gauge coupling $g_X$, even when these additional annihilation channels 
are considered.

\subsection{Local $Z_2$ fermion DM: $Q_\chi=1, Q_\Phi =2$}

In the context of the Dirac fermion DM model, we explore a specific scenario characterized by $Q_\chi=1, Q_\Phi =2$.
This choice modifies the DM Lagrangian at the renormalizable level:
\begin{equation}
\mathcal{L}_{\rm DM} =\overline{\chi} ( i \slashed{D} - m{\chi} ) \chi - \left( y_\Phi \overline{\chi^C} \chi \Phi^\dagger + \text{H.c.} \right).
\end{equation}

The symmetry breaking pattern is ${\rm U}(1)_X \to Z_2$ due to the $y_\Phi$ terms. This model, serving as a dark gauge model for inelastic fermion DM, has been previously investigated in the context of DM bound state formation \cite{Ko:2019wxq} and the XENON1T excess \cite{Baek:2020owl}. A comprehensive study of the light dark Higgs contribution to the DM self-interaction and relic density is presented in \cite{Kamada:2018zxi}.

Upon ${\rm U}(1)_X$ symmetry breaking with \added{a} nonzero $v_\Phi$, the original Dirac fermion $\chi$ decomposes into two Majorana fermions ($\chi_R$ and $\chi_I$) with a mass splitting proportional to $v_\Phi$:
\begin{equation}
\delta \equiv M_{R} - M_{I} = 2y_\Phi v_\Phi.
\end{equation}

Assuming $y_\Phi>0$, we ensure $\delta >0$, causing the lighter state $\chi_I$ to become Majorana fermion DM candidate, 
while $\chi_R$ becomes its excited state. The Lagrangian for them is expressed as follows:
\begin{align}
\mathcal{L}_{\rm DM} &= \frac{1}{2} \sum_{i=I,R} \bar{\chi}_i \left( i\partial_\mu \gamma^\mu - M_i \right) \chi_i -i \frac{g_X}{2} Z'_\mu \left( \bar{\chi}_R \gamma^\mu \chi_I  - \bar{\chi}_I \gamma^\mu \chi_R \right) 
\nonumber \\ 
& -\frac{1}{2} y_\Phi \left( c_\alpha H_1 + s_\alpha H_2  \right) \left( \bar{\chi}_R \chi_R - \bar{\chi}_I \chi_I \right).
\end{align}
We analyze the thermal relic density within the standard freeze-out mechanism. Notably, the singlet scalar $\Phi$ plays a crucial role in enabling the fermionic DM to be a thermal WIMP, allowing for DM masses up to $\sim O({\rm a~ few})$ TeV. 
This result is in sharp contrast with the results found in the literature where the $Z'$ 
mass is assumed to be generated  without the 
dark Higgs boson by St\"{u}ckelberg mechanism.

\begin{figure}
\centering
\includegraphics[width=0.45\linewidth]{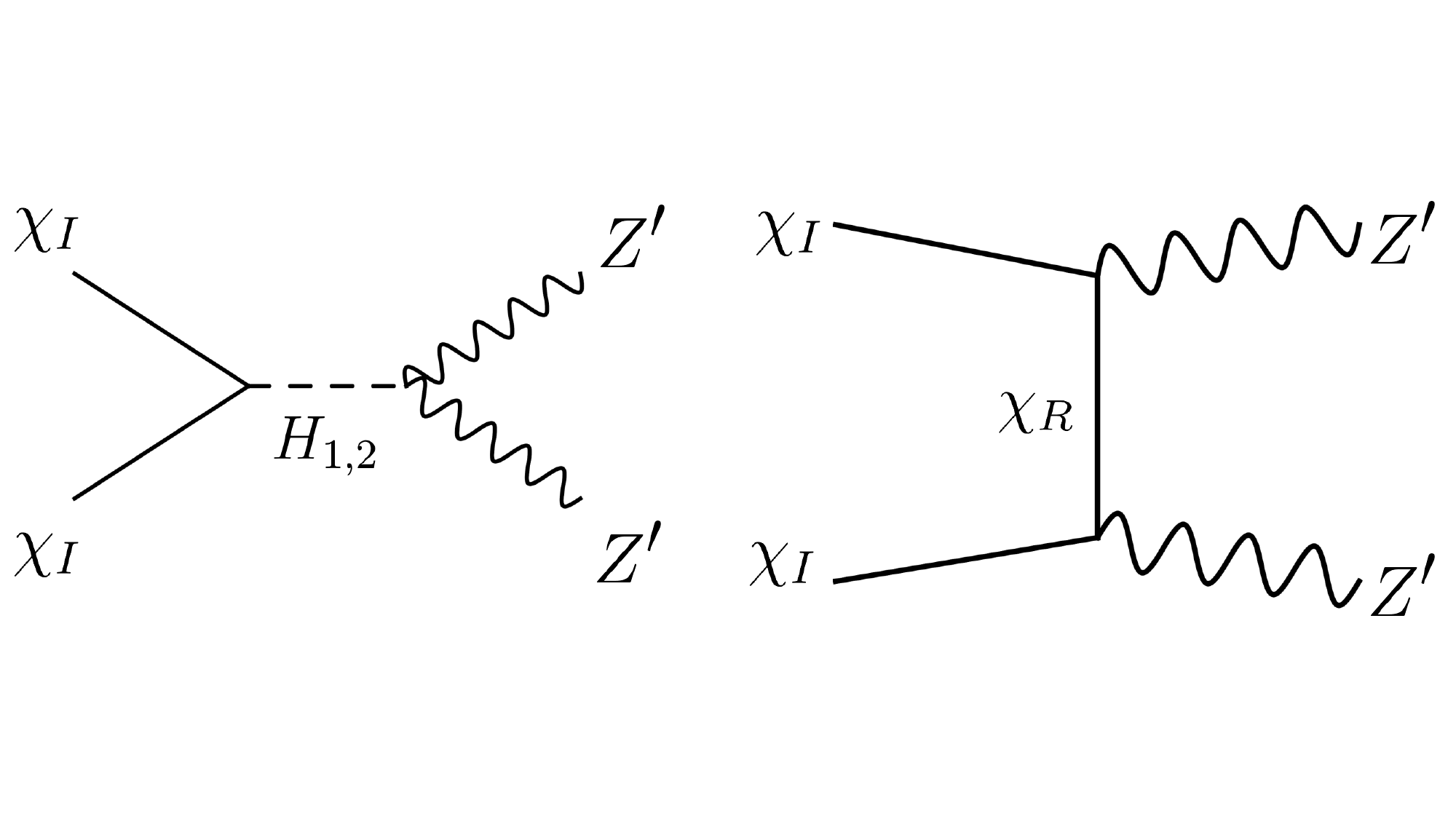}
\includegraphics[width=0.45\linewidth]{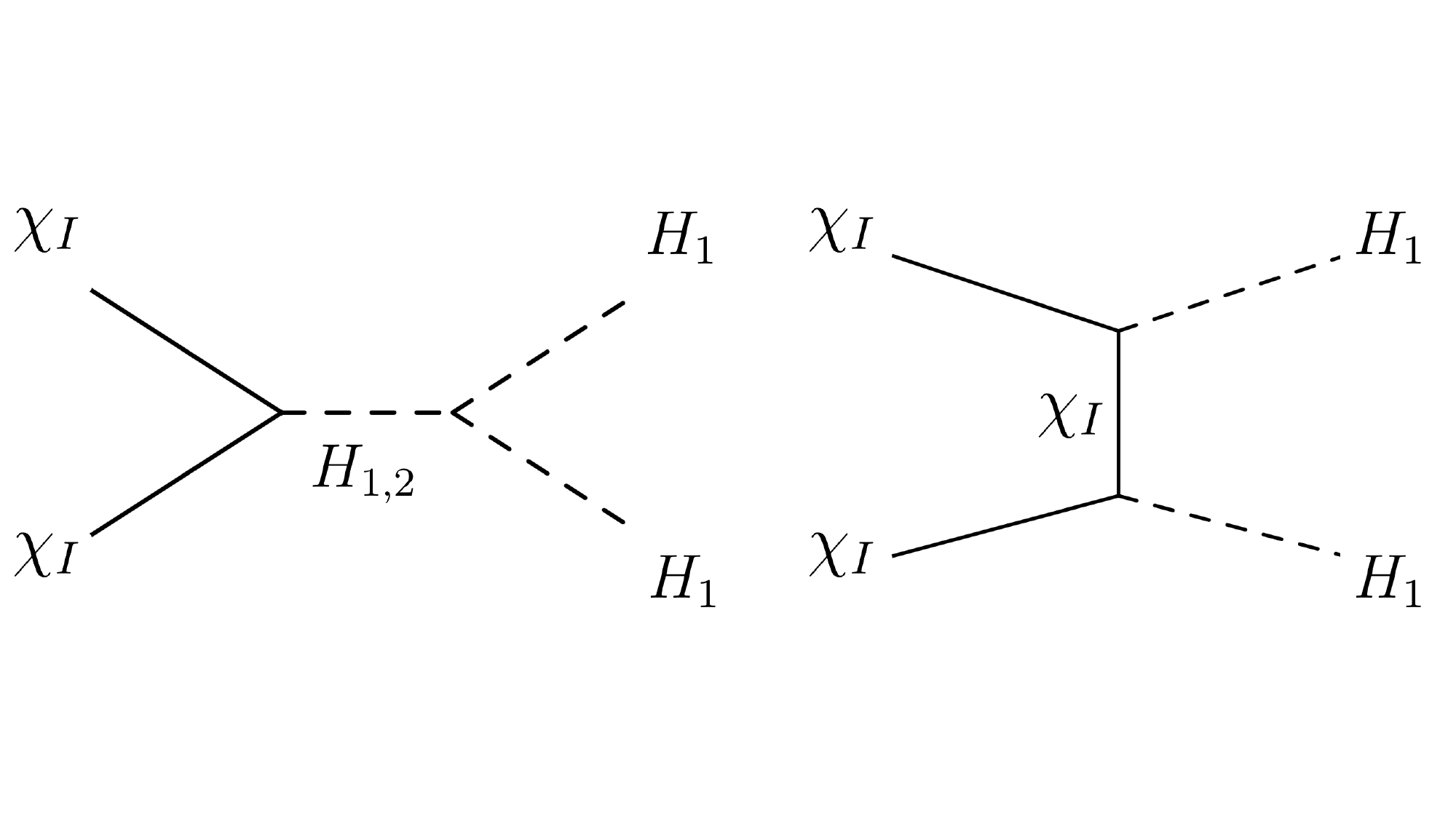}
\hspace{0.5cm}
\includegraphics[width=0.22\linewidth]{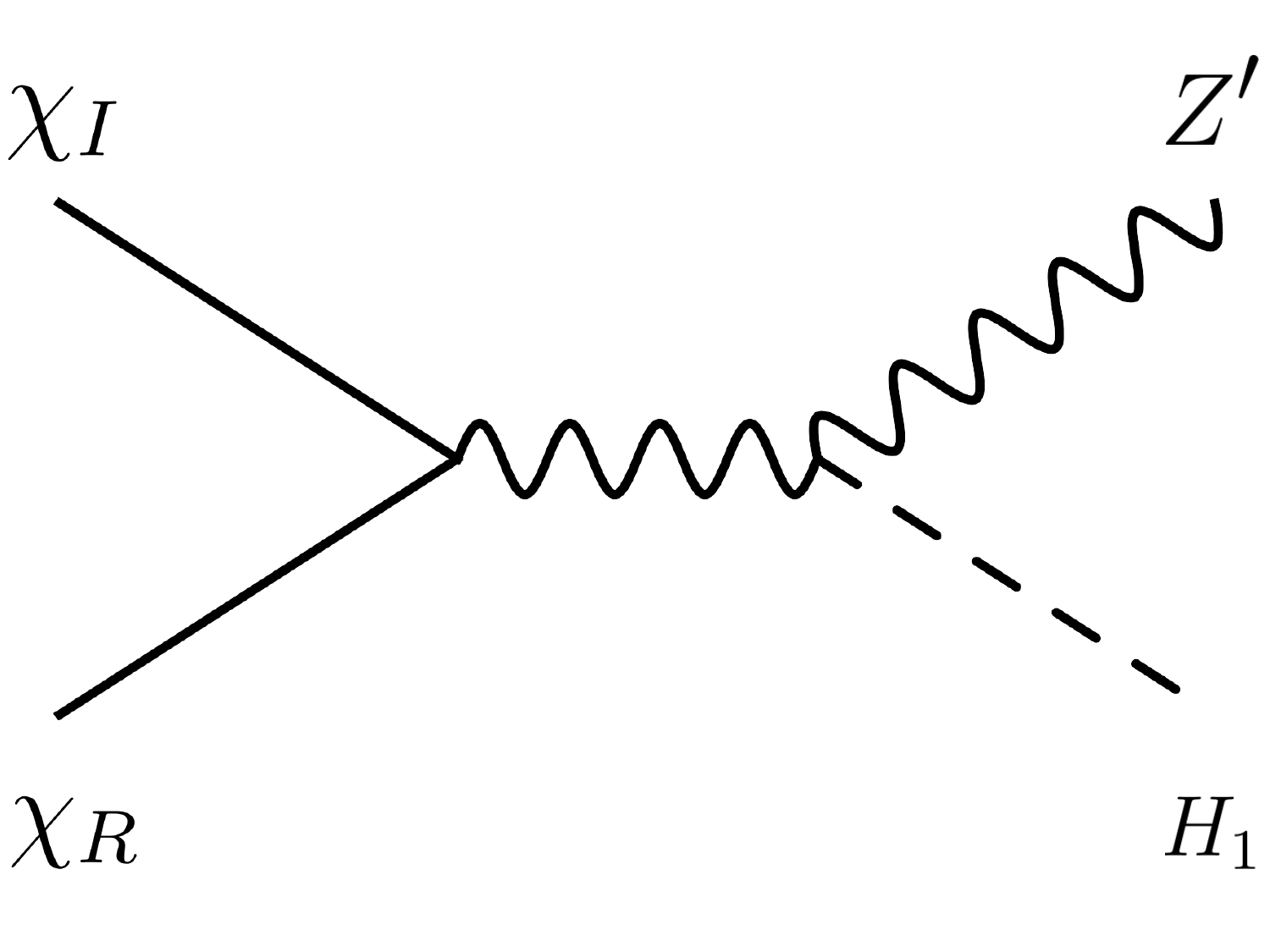}
\includegraphics[width=0.22\linewidth]{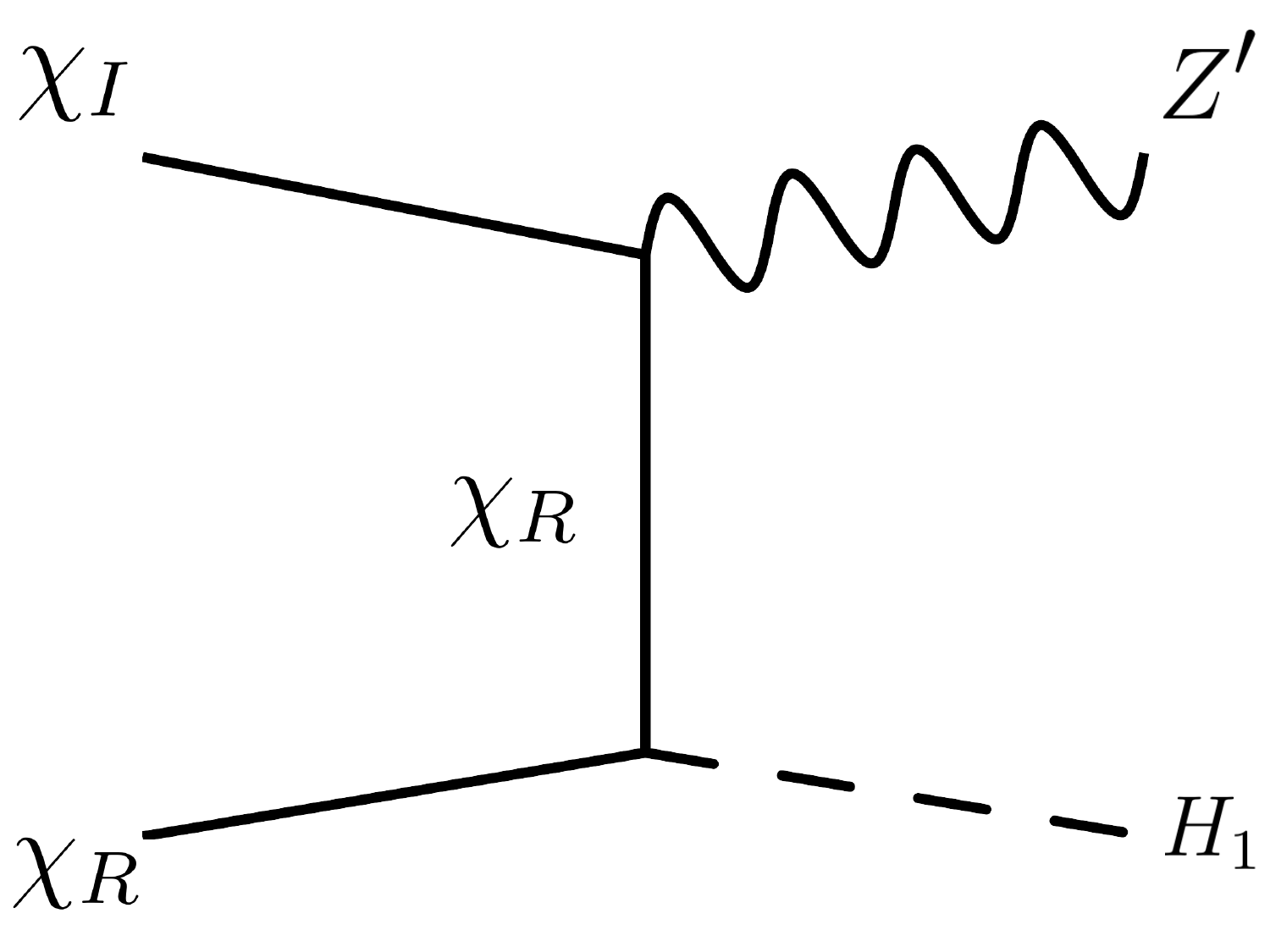}
\includegraphics[width=0.22\linewidth]{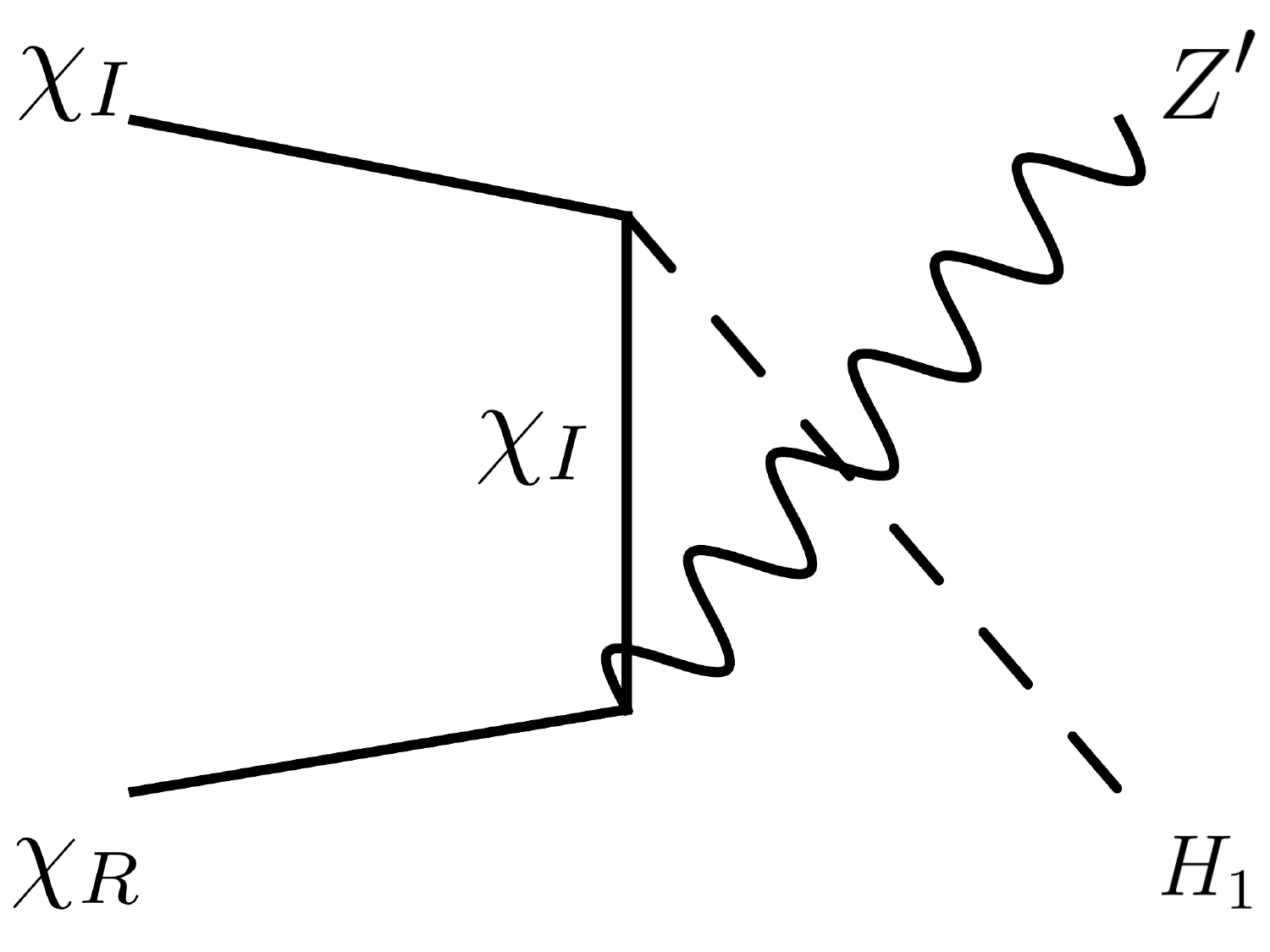}
\hspace{0.5cm}
\caption{Feynman diagrams of local $Z_2$ fermion DM (co-)annihilating into a pair of $Z'$ bosons and $H_1$ bosons  ({\it Top}), and $Z' + H_1$ ({\it Bottom}). \label{FeynmanZ2FDM}} 
\end{figure}
Particularly, the (co-)annihilation channel involving $H_1$ significantly influences the determination of the relic density, a channel often omitted in conventional approaches. 
The corresponding Feynman diagrams are described in Fig.~\ref{FeynmanZ2FDM}.

\begin{figure}
\centering
\includegraphics[width=0.45\linewidth]{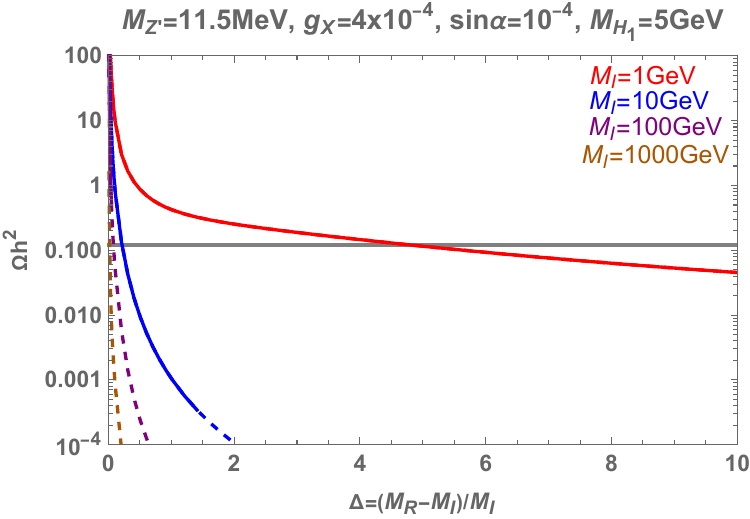}
\includegraphics[width=0.45\linewidth]{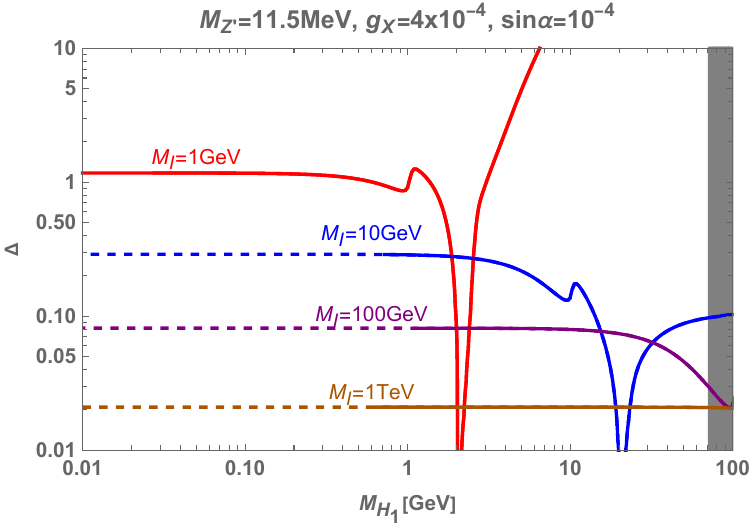}
\hspace{0.5cm}
\caption{Plots for the local $Z_2$ fermionic DM model. {\it Left}:  Dark matter relic density as a function of mass splitting $\Delta$ for [{\bf BPI}] and for different values of DM mass, $M_I = 1,10,100, 1000$GeV. Solid (Dashed) lines denote the region where bounds on DM direct detection are satisfied (ruled out).
{\it Right}: Preferred parameter space in the $(M_{H_1} , \Delta)$-plane for  different DM masses. The gray region is ruled out by the perturbativity condition on $\lambda_\Phi$.  \label{localZ2FDM}}
\end{figure}

In Fig.~\ref{localZ2FDM}, we show thermal relic density plots.
On the left of Fig.~\ref{localZ2FDM}, we depict $\Omega_{\rm DM}h^2$ as a function of the mass splitting parameter $\Delta$ for the [{\bf BPI}] scenario, 
considering $M_{H_1} = 5$ GeV and varying DM masses $M_I = 1, 10, 10^2, 10^3$ GeV. 
The most significant contributions originate from $\chi_I \chi_I \to Z'Z', H_1H_1$ and $\chi_I \chi_R \to H_1 Z'$. 
Solid (Dashed) lines delineate the region satisfying (excluded by) DM direct detection bounds.
In the limit as $\Delta$ approaches zero, the relic density is found to be excessively large irrespective of the selected dark matter masses, 
thus affirming the assertion made in Section \ref{sec:fDM_generic} that the relation expressed in Eq.~(\ref{eq:res}) is indispensable for generic fermionic dark matter. 
Furthermore, it becomes evident that smaller mass splittings $\Delta$ are necessary for heavy fermionic dark matter in order to meet the conditions for relic abundance.
For the [{\bf BPI}] scenario, co-annihilation 
becomes important when DM is heavy and the mass splitting is small, due to the $t(u)$-channel diagrams of the co-annihilation process involving a $y_\Phi$ coupling. 
The $(M_{H_1}, \Delta)$-plane for the same choices of $M_I$ is shown. 
The right side of Fig.\ref{localZ2FDM} displays contours of $\Omega h^2 = 0.12$.

\begin{figure}
\centering	
\includegraphics[width=0.45\linewidth]{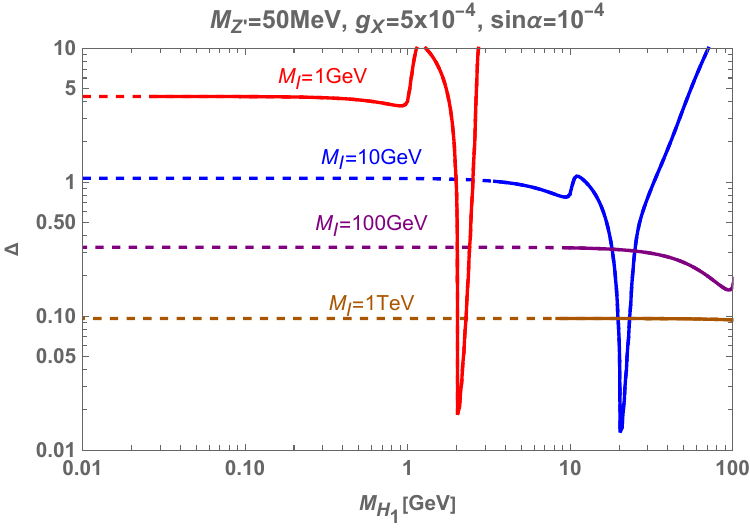}	
\includegraphics[width=0.45\linewidth]{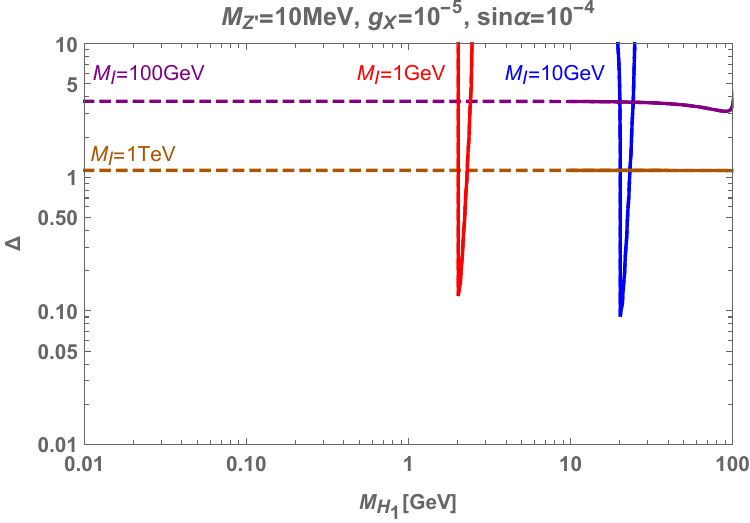}	
\caption{
Plots corresponding to Fig.~\ref{localZ2FDM} for [{\bf BPII}] (left) and [{\bf BPIII}] (right).
}	
\label{localZ2FDMBP2}
\end{figure}

Similar plots for the [{\bf BPII}] and [{\bf BPIII}] scenarios are presented in Fig.~\ref{localZ2FDMBP2}.
In contrast to the [{\bf BPI}], for [{\bf BPII}], co-annihilation is less significant for obtaining the correct $\Omega h^2$, 
given the large value of $v_\Phi$. 
For [{\bf BPIII}], we have large $v_\Phi$ compared to  [{\bf BPI}] and  [{\bf BPII}]. 
Therefore, all the lines move up.

The process of DM direct detection in the local $Z_2$ fermion DM case is akin to the local $Z_2$ scalar 
DM case.  Again, there are two processes:
elastic scattering mediated by (dark) Higgs exchanges and inelastic scattering through $Z'$ exchange with kinetic mixing. 
However, for  $\delta \gtrsim {\cal O}(10)$ keV, the inelastic scattering process is negligible. 
Consequently, our focus is solely on the Higgs-mediated elastic scattering, the explicit form of which is given by
\begin{eqnarray}
\sigma_{\rm SI} &=& \frac{\mu^2_N}{\pi} \Delta^2 \left( \frac{ M_I M_N}{v_H v_\Phi} \right)^2 f^2_N s^2_\alpha c^2_\alpha \left( \frac{1}{M^2_{H_1}} - \frac{1}{M^2_{H_2}}\right)^2,
\end{eqnarray}
where $\Delta \equiv \left( M_R - M_I \right)/M_I$.

Imposing this constraint, the resulting plots in Fig.~\ref{localZ2FDM} reveal the dashed regions excluded by DM direct detection experiments. 
Nevertheless, a substantial parameter space exists for heavy DM masses $M_I$, well beyond $M_I \sim M_{Z'}/2 \sim O(10-100)$ MeV.

\section{Conclusions}
In this study, we investigated the implications of a light $Z'$ on thermal dark matter, 
incorporating both scalar and Dirac fermion dark matter (DM) candidates within the framework of 
$U(1)_{L_\mu-L_\tau}$ extensions to the Standard Model (SM). 
The light $Z'$  is not only sensitive to the muon anomalous magnetic moment $(g-2)_\mu$, but it can also significantly enhance the dark matter annihilation cross section through its longitudinal polarization. This enhancement is most relevant in the $X X^\dagger \to H_1 \to Z'_L Z'_L$ channel.
Traditionally, thermal DM is introduced under the assumption that the dark matter mass meets the condition $M_{\text{DM}} \sim M_{Z'}/2$. 
The correct DM relic density is achieved by DM annihilation through the $Z'$ resonance, which dictates $M_{\text{DM}} \sim M_{Z'}/2 \approx \mathcal{O}(10)$ MeV.

In this paper, we considered the contributions of the dark Higgs boson, an often overlooked aspect in previous studies, in various DM models. 
Notably, we found that the stringent mass relationship $M_{\text{DM}} \sim M_{Z'}/2$ can be circumvented.
Our analysis  demonstrated that weakly interacting massive particles (WIMPs) with masses ranging from the GeV to TeV scales, significantly 
exceeding the $M_{Z'}\approx \mathcal{O}(10)$ MeV constraint indicated by the above mass relationship
in the absence of the dark Higgs boson, can still be viable thermal WIMP DM candidates without violating other constraints. 
We have established that the dark Higgs is essential for achieving the electroweak-scale dark matter. With the dark Higgs in place, DM annihilation into light $Z'$ modes is enhanced, facilitating efficient thermal DM production.
Note that these solutions are independent of the $Z'$ contribution to $\Delta a_\mu$, as  DM annihilations are primarily controlled by the dark matter and the dark Higgs coupling $\lambda_{\Phi X}$, and the dark Higgs mass  $M_{H_1}$, while $\Delta a_\mu$ is governed by the dark gauge coupling $g_X$.

The details of DM phenomenology depend on the $U(1)$ charge assignments to the DM and the dark Higgs ($\Phi$) fields. The $U(1)_{L_\mu-L_\tau}$ symmetry can be broken in a generic  or a specific manner, leading to either $Z_2$ (inelastic scalar or fermion DM models) or a $Z_3$ scalar DM model. The inclusion of the dark Higgs boson introduces new DM (co-)annihilation channels, such as ${\rm DM}+{\rm \overline{DM}} \to H_1 H_1, H_1 Z'$, and ${\rm DM}+{\rm \overline{DM}} \to H_1 \to Z' Z'$. In the latter process, there is an enhancement in longitudinal $Z'$ pair production. These newly opened channels significantly broaden the viable DM mass range, spanning from the GeV to a few TeV 
\footnote{Recently two of the present authors also explored lighter DM $(X)$ mass regions 
($O(10)$ MeV $\lesssim m_X \lesssim 10$ GeV) 
in the same gauge models \cite{Ho:2024cwk}, motivated by the Belle II announcement of the branching ratio for  
$B^+ \rightarrow K^+ \nu \bar{\nu}$, which was reported to exceed the SM prediction by  $2.7 \sigma$. 
In Ref. \cite{Ho:2024cwk},  the excess was interpreted as the $B^+ \rightarrow K^+ +$ dark sector particles,  
either into DM pair or dark photon pair. In the latter case,  the dark photon decays invisibly into a pair of
the SM neutrinos,  $Z' \rightarrow \nu \bar{\nu}$.}, 
while satisfying the constraints from the direct DM searches.
This deviation from the traditional tight correlation between $M_{Z'}$ and $M_{\rm DM}$ ($M_{\rm DM} \sim M_{Z'}/2$) underscores the importance of considering the dark Higgs boson for a comprehensive understanding of DM phenomenology in the presence of a massive dark photon. Our analysis clearly demonstrates that a thorough exploration of DM phenomenology involving a massive dark photon necessitates the inclusion of the dark Higgs boson.

The models under consideration offer opportunities for further tests through dark matter indirect detection and collider experiments. 
Among the most significant DM annihilation channels are ${\rm DM}+{\rm \overline{DM}} \to H_1 H_1, H_1 Z', Z' Z'$. 
The resulting $H_1$ predominantly undergoes decay into a pair of $Z'$ bosons. In our analysis, we have adopted benchmark points with  $M_{Z'} = 10$ MeV, $11.5$ MeV and $50$ MeV.
Consequently, for $M_{Z'} < M_\mu$, $Z'$ decays into either muon-type or tau-type neutrinos.
Anticipating DM annihilation in the vicinity of the Galactic center, we predict a substantial neutrino flux. 
This flux stemming from DM annihilation stands as a target for detection by experiments like IceCube and Hyper-Kamiokande detectors \cite{Baek:2008nz, Asai:2020qlp}. 
As the dark photon exclusively couples to the second- and third-generation leptons, 
the proposed muon colliders represent the most promising avenue for probing the model. 
However, verifying $U(1)_{L_\mu-L_\tau}$-charged DM models at proton-proton colliders, such as the LHC, poses challenges. 
Nevertheless, investigations near the $Z$ boson mass region can be pursued through processes like $pp\to Z\to 2\mu Z' \to 4\mu$ 
as outlined in~\cite{CMS:2018yxg}, even though the allowed zone conflicts with the current muon $(g-2)$ data.

\appendix \label{appendix}
\section{Constraints on the ${\rm U(1)}_{L_\mu - L_\tau}$ Models}
In this Appendix we collect various constraints imposed on our model.
\subsection{Muon $(g-2)$}

The magnetic moment of the muon is expressed as
\begin{eqnarray}
\vec{\mu}_\mu &=& g_\mu \left( \frac{e}{2m} \right) \vec{S},
\end{eqnarray}
where $g_\mu$ denotes the gyromagnetic ratio. Considering radiative corrections, the anomalous magnetic moment of the muon is defined by
\begin{eqnarray}
a_\mu &=& \frac{1}{2} \left( g_\mu -2 \right).
\end{eqnarray}

In the ${\rm U(1)}_{L_\mu - L_\tau}$ models, the additional contribution of the new gauge boson $Z'$ to the muon $(g-2)$ arises through the vertex correction to the muon-photon coupling \cite{Baek:2001kca,Baek:2008nz}:
\begin{equation}
\Delta a_\mu = \frac{\alpha_X}{2 \pi} \int_0^1 dx \frac{2 M_\mu^2 x^2 (1-x)}{x^2 M_\mu^2 + (1-x) M_{Z'}^2},
\end{equation}
where $\alpha_X=g^2_X/4\pi$.

If one extends the particle content by including exotic leptons and/or dark matter,
additional contributions from heavier $Z'$ become relevant, providing the potential to accommodate $\Delta a_\mu$. This aspect has been discussed in the literature \cite{Borah:2021mri, Qi:2021rhh, Singirala:2021gok, Buras:2021btx, Zhou:2021vnf, Borah:2021jzu, Chen:2021vzk, Zu:2021odn, Huang:2021nkl, Patra:2016shz, Altmannshofer:2016oaq}.
\subsection{BaBar and LHC 4$\mu$ channels}
Dark photon searches at electron beam dump experiments impose constraints on our model's parameter space. For instance, BaBar conducted searches for the process $e^+ e^- \to Z' \mu^+ \mu^-$ followed by $Z' \rightarrow \mu^+ \mu^-$ \cite{BaBar:2016sci}. While the BaBar bounds depend on the branching ratio of $Z'$ decaying to DM particles, for our mass range of interest, $M_{Z'} \approx 10-100$ MeV, this channel is kinematically forbidden.
We use the $2\sigma$ exclusion limit from the BaBar bounds, as depicted in Fig.~1.

The $Z\to 4\mu$ measurements at the LHC also provide stringent constraints on our models. The LHC has observed $pp \to Z \to 4\mu$ channels, where four muon events near an invariant mass close to the $Z$ boson mass are analyzed. In the context of the $U(1)_{L_\mu - L_\tau}$ model, the produced $Z$ boson might decay to $Z'$ and a muon pair. If $M_{Z'} > 2 M_\mu$, the produced $Z'$ subsequently decays to a pair of muons. Therefore, these events are sensitive to the $pp \to Z \to 4\mu$ decay channel. We consider the $2\sigma$ exclusion limit from the CMS data \cite{CMS:2018yxg}.
\subsection{Neutrino trident production}
Neutrino beam experiments, such as CHARM-II \cite{CHARM-II:1990dvf} and CCFR \cite{CCFR:1991lpl}, have established neutrino trident production, wherein muon-neutrino scattering on a nucleus produces a muon pair, $\nu_\mu N \to \nu_\mu N \mu^+ \mu^-$ \cite{Altmannshofer:2014pba}. The observed scattering cross sections are
\begin{align}
\frac{\sigma_{\rm CHARM-II}}{\sigma_{\rm SM}} = 1.58 \pm 0.57,\qquad \frac{\sigma_{\rm CCFR}}{\sigma_{\rm SM}} = 0.82 \pm 0.28.
\end{align}
The observational data is consistent with the Standard Model (SM) prediction. Therefore, neutrino trident production imposes a stringent constraint, essentially excluding the parameter region with $M_{Z'} > O(1)$ GeV \cite{Altmannshofer:2014pba}. In our study, we consider the $2\sigma$ exclusion limit from the CCFR data.

\subsection{Cosmic Microwave Background }

There is an interaction between $Z'$ and neutrinos in our model. We assume that $Z'$ and DM were in thermal equilibrium with the SM bath in the early Universe. If the $Z'$ mass is much below the muon mass, the $Z'$ decay process will provide energy and entropy to neutrinos. If this occurs after the epoch of neutrino decoupling, corresponding to $T\sim O(1)$ MeV, it introduces additional entropy and energy, potentially affecting the successful Big Bang Nucleosynthesis (BBN). This effect modifies $N_{\rm eff}$, and we consider $N_{\rm eff} < 3.4$ as the exclusion limit.

However, there exists a discrepancy between the observed Hubble constant today, as derived from the Cosmic Microwave Background, and other astrophysical observations. This is known as the Hubble tension. To address the Hubble tension, an increase in $N_{\rm eff}$ is considered beneficial \cite{Escudero:2019gzq, Huang:2021dba, DiValentino:2021izs}.
For two of our benchmark points, {\bf BPIII} and {\bf BPI}, $(M_{Z'}, g_X)= (10 \text{ MeV}, 10^{-5})~ \text{and} ~(11.5 \text{ MeV}, 4\times 10^{-4})$, we allow $N_{\rm eff} > 3.2$ to reconcile this Hubble tension.

\subsection{Borexino}
Reactor neutrino experiments can offer valuable constraints on $g_X$ and $M_{Z'}$ through the neutrino-electron scattering process. The Borexino collaboration provides highly sensitive data \cite{Bellini:2011rx} on this channel. The $^7{\rm Be}$ neutrinos, $\nu_e$, undergo oscillations to $\nu_\mu$ and $\nu_\tau$ on their way to Earth. At the detector, they scatter off electrons via $Z'$ exchange \cite{Harnik:2012ni,Bauer:2018onh}. The corresponding bound is presented in Fig.~1.

\subsection{NA64}

The NA64 experiments utilize a high-energy muon beam and a missing energy-momentum technique to search for light $Z'$~\cite{Andreev:2024sgn}. 
In our model, the event process is $\mu N \to \mu N Z'$, followed by $Z' \to \bar{\nu} \nu$. 
NA64 results exclude a large part of parameter space of $(M_{Z'}, g_X)$.

\subsection{Dark Higgs Decay}
The decay width of the dark Higgs to a lepton pair is given by
\begin{eqnarray}
\Gamma\left( H_1 \to \ell^+ \ell^-  \right) &=& \frac{M^2_\ell M_{H_1} }{8\pi v^2_H} \sin^2\alpha \left(1- \frac{4M^2_\ell}{M^2_{H_1}} \right)^{3/2},
\end{eqnarray}
where $\ell=e, \mu,\tau$.
If kinematically allowed, the dark Higgs can decay to a pair of $Z'$ bosons, and the corresponding width is given by
\begin{eqnarray}
\Gamma \left(H_1 \to Z'Z'  \right) &=& \frac{g^2_X c^2_\alpha}{32\pi} \frac{M^3_{H_1}}{M^2_{Z'}} \left( 1 - \frac{4M^2_{Z'} }{M^2_{H_1}} + \frac{12M^4_{Z'} }{M^4_{H_1}}\right).
\end{eqnarray}
For large $\lambda_{\Phi X}$ and $M_{H_1} \geq 2 M_X$, the dark Higgs predominantly decays to a pair of DMs which is given by
\begin{align}
\Gamma \left(H_1 \to XX^\dagger \right) &\simeq \frac{\lambda^2_{\Phi X} v^2_\Phi \cos^2\alpha}{ 16\pi M_{H_1}} \sqrt{1-\frac{4 M^2_X}{M^2_{H_1}} }.
\end{align}
In our parameter space of interest, corresponding to $M_{Z'} \sim \mathcal{O}(10)$ MeV and $M_{H_1} \geq \mathcal{O}(0.1)$ GeV, the dark Higgs predominantly decays into a pair of either $Z'$ bosons or DMs.
$Z'$ subsequently decays to $\nu_\alpha \overline{\nu}_\alpha$ with $\alpha = \mu , \tau$. 
The produced neutrinos might shift $\Delta N_{\rm eff}$.
However, this effect is negligible since the comoving number density $Y_{H_1}$ 
at the decoupling epoch is very small.
For $M_{H_1} \leq 2 M_{Z'}$, we require the lifetime of the dark Higgs to be less than $\sim O(1)$ sec to avoid the bound from BBN.

\acknowledgments
This work is in part supported by KIAS Individual Grants, Grant No. PG021403 (PK) 
and Grant No. PG074202 (JK) at Korea Institute for Advanced Study, and by National Research Foundation of Korea (NRF) Grant No. NRF-2018R1A2A3075605, No. RS-2023-00270569 (SB), 
No. NRF-2022R1A2C2003567, No. RS-2024-00341419 (JK),
and No. NRF-2019R1A2C3005009 (PK), funded by the Korea government (MSIT).

\bibliographystyle{JHEP}
\bibliography{biblio.bib}

\end{document}